\documentclass[a4paper,11pt]{article}
\pdfoutput=1 

\usepackage{jheppub} 

\usepackage{comment}
\usepackage{amsmath,amssymb,mathtools,graphicx,setspace}
\usepackage{slashed} 		 
\usepackage{color}
\usepackage{caption}
\usepackage{subcaption}
\makeatother

\newcommand{\be}{\begin{equation}}
\newcommand{\bea}{\begin{eqnarray}}
\newcommand{\eea}{\end{eqnarray}}
\newcommand{\ba}{\begin{array}}
\newcommand{\ea}{\end{array}}
\newcommand{\ee}{\end{equation}}
\newcommand{\bes}{\begin{equation*}}
\newcommand{\beas}{\begin{eqnarray*}}
\newcommand{\eeas}{\end{eqnarray*}}
\newcommand{\bas}{\begin{array*}}
\newcommand{\eas}{\end{array*}}
\newcommand{\ees}{\end{equation*}}

\numberwithin{equation}{section}

\title{Complexity via Replica Trick}

\author[a]{Mohsen Alishahiha,}
\emailAdd{alishah@ipm.ir}
\affiliation[a]{School of Physics, Institute for Research in Fundamental Sciences (IPM),\\
	P.O. Box 19395-5531, Tehran, Iran\\} 
\author[b,c]{Souvik Banerjee,}
\emailAdd{souvik.banerjee@physik.uni-wuerzburg.de}
\affiliation[b]{Institut f{\"u}r Theoretische Physik und Astrophysik,
	Julius-Maximilians-Universit{\"a}t W{\"u}rzburg,\\ Am Hubland, 97074 W{\"u}rzburg, Germany\\} 
\affiliation[c]{W\"urzburg-Dresden Cluster of Excellence ct.qmat\\}
\author[d]{Joshua Kames-King}
\emailAdd{jvakk@yahoo.com}
\affiliation[d]{Bethe Center for Theoretical Physics 
and
Physikalisches Institut der Universitaet Bonn, Nussallee 12, 53115 Bonn, Germany\\}

\abstract{We consider the complexity of a single-sided AdS black hole as modelled by an end-of-the-world brane. In addition we present multi-boundary partition functions and matter correlation functions for such a setting. We compute the complexity using a modified replica trick corresponding to the ``quenched geodesic length" in JT gravity. The late time behaviour of complexity shows a saturation to a constant value of order $e^{S_0}$ following a period of linear growth. Furthermore, we show that our approach leads to an improved result for the variance of complexity, namely it being time-independent  at late times. We conclude by commenting on the introduction of dynamical end-of-the-world branes.}

\begin{document} 
\maketitle
\flushbottom

\section{Introduction}
\label{sec:intro}

In the context of the AdS/CFT correspondence 
\cite{Maldacena:1997re, Witten:1998qj,Gubser:1998bc}
it is believed that the interior of a black hole may be systematically studied via the notion of 
quantum computational complexity. This field of study quantifies the difficulty of constructing a 
specific  ``target state" by use of a simple set of ``universal gates". More specifically in a 
holographic setting it is conjectured that for a chaotic CFT the growth of complexity has a simple 
geometric description in terms of the growth of the black hole interior.

One of the arguments for this conjecture is that for a fast-scrambling system with finite entropy $S
$, complexity is expected to grow for exponentially large times in the entropy, long after thermal 
equilibrium has been reached \cite{Susskind:2014rva,Susskind:2018pmk}. Remarkably, the same 
growth holds for the black hole interior. Therefore a concrete instantiation of this conjectured 
duality is the ``complexity=volume" (CV) conjecture, which proposes that the complexity equals 
the volume of a maximal slice in the black hole interior \cite{Susskind:2014rva, Stanford:2014jda}. 
There is also another competing proposal known as the ``complexity=action" (CA) conjecture, 
in which the on-shell action on a Wheeler-de Witt patch is determined 
\cite{Brown:2015bva,Brown:2015lvg}.
   
We note however, that for chaotic Hamiltonians (as can, for example, be seen in simple circuit 
models) after the aforementioned period of growth, at times $t \sim \left( \mathcal{O}(e^S)\right)$ 
we expect saturation to a plateau of size $C \sim \left( \mathcal{O}(e^S)\right)$ 
\cite{Susskind:2015toa,Brown:2016wib,Balasubramanian:2019wgd,
Susskind:2020wwe,Balasubramanian:2021mxo,Haferkamp:2021uxo}. 
While semi-classical contributions both in form of the CV and CA 
conjectures indeed furnish the period of growth, the saturation to the plateau, until recently, has 
been illusive.

To understand complexity better one may study this concept in Jackiw-Teitelboim (JT) gravity; a theory of two-dimensional dilaton gravity, including arbitrary 
genus, hyperbolic Riemann surfaces and therefore also exponentially small corrections to semi-classical gravity calculations \cite{Teitelboim:1983ux,Jackiw:1984je,
Maldacena:2016upp,Engelsoy:2016xyb,Saad:2018bqo, Saad:2019lba}.
 Actually extending the gravitational 
sector by including such geometries with an arbitrary number of asymptotic boundaries and 
arbitrary genus corrects the partition function to be equivalent to a specific double-scaled 
Hermitian matrix integral. This implies that JT gravity follows RMT universality at late times and 
therefore exhibits spectral statistics with a dip-ramp-plateau structure 
\cite{Saad:2018bqo,Saad:2019lba,Saad:2019pqd,Altland:2020ccq}.\footnote{For work 
on the relationship between chaos  universality and Euclidean wormholes in higher 
dimensions see \cite{Belin:2020hea,Cotler:2020ugk}.} By use of the same theory it has 
also been shown that the inclusion of higher 
topologies gives a unitary Page curve \cite{Almheiri:2019qdq,Penington:2019kki}.

 Recently, holographic  complexity was calculated in JT gravity using the CV conjecture in \cite{Iliesiu:2021ari} where it was shown that  including  higher genus geometries 
 (as mentioned above) gives the correct late-time behaviour for complexity. 
More precisely, in this paper the authors compute complexity in terms of a
non-perturbative geodesic length  in JT gravity as follows
\begin{equation}\label{eq:nonperturbativelengthintroduction}
    \langle \ell \rangle=\lim_{\Delta \rightarrow 0} \left\langle \sum_{\gamma} \ell_{\gamma} e^{-
  \Delta \ell_{\gamma}} \right\rangle_{\text{JT}}\,,
\end{equation}
where $\gamma$ refers to non self-intersecting geodesics, $\Delta$ is a regulator and 
$\langle \rangle_{\text{JT}}$ a correlator in JT gravity defined over arbitrary genus. 
It is then argued that in practice \eqref{eq:nonperturbativelengthintroduction} is calculated by
\begin{equation}\label{eq:nonperturbativetimedependentlengthintroduction}
    \langle \ell (t)\rangle=-\lim_{\Delta \rightarrow 0} \frac{\partial \langle \chi(t) 
    \chi(0)\rangle_{\text{non-int.}}}{\partial \Delta}\,,
\end{equation}
where $\langle \chi(t) \chi(0)\rangle_{\text{non-int.}}$  is obtained in the Euclidean JT theory and 
then analytically continued. Here $\Delta$ is the scaling dimension of the operator $\chi$. Eq. 
\eqref{eq:nonperturbativetimedependentlengthintroduction} then of course involves (on surfaces 
with $g\geq 1$) an infinite number of geodesics which can be taken care of by the moduli space 
volume of hyperbolic surfaces \cite{Mirzakhani:2006fta}. 
It was, then,  demonstrated that the above definition results in the following 
expression for complexity
\begin{equation}\label{eq:spectralcomplexityintroduction}
    \langle \ell(t) \rangle=-\frac{2 e^{-S_0}}{Z(\beta)}\int_{0}^{\infty}\frac{\langle \rho(s_1) 
    \rho(s_2)\rangle}{\bar{s}\sinh (2 \pi \bar{s}) \omega \sinh (\pi \omega)}\exp\left(-\beta
    \left(\frac{\bar{s}^2}{2}+\frac{\omega^2}{8}\right)-i\bar{s}\omega t\right)\,.
\end{equation}
with the definitions of $\omega=s_1-s_2\,,\bar{s}=\frac{s_1+s_2}{2}$ and 
$s_{1,2}=\sqrt{E_{1,2}}\;$. The quantity \eqref{eq:spectralcomplexityintroduction} was called ``spectral complexity" in 
\cite{Iliesiu:2021ari},  which can be calculated for any quantum theory by use of
its spectral correlation $\langle \rho(s_1) \rho(s_2)\rangle$.

Due to the usual arguments regarding quantum chaos 
\cite{PhysRevA.43.2046,PhysRevE.50.888}, one would suspect that for chaotic systems,
\eqref{eq:spectralcomplexityintroduction} would reduce to RMT predictions at late times. For the 
case of JT gravity, the spectral two-point function can be shown to take on the standard RMT sine-kernel structure \cite{Saad:2019lba,Okuyama:2020ncd,Altland:2020ccq} by use of doubly non-perturbative effects. This in turn leads to the aforementioned, expected behaviour for the quantity 
$\langle \ell(t)\rangle$: early linear growth followed by a late-time plateau saturation.

 In the present  work, we are interested in studying two aspects of complexity for JT gravity. First, we would like to use an approach which removes the worrisome behaviour of the variance obtained in \cite{Iliesiu:2021ari}, as we will explain in greater detail below. Secondly, we would like to study the introduction of an end-of-the-world (EOW) brane. Recently, these objects have played a crucial role in
understanding quantum aspects of black holes in a two-dimensional setting as they can be used 
to model black hole microstates in JT gravity \cite{Penington:2019kki}.
Since a black hole with an EOW brane behind the horizon may be understood as a 
$\mathcal{Z}_2$ quotient of the two-sided scenario, it corresponds to a pure state 
\cite{Kourkoulou:2017zaj,Maldacena:2001kr}. However, according to the eigenstate thermalisation 
hypothesis (ETH) \cite{PhysRevA.43.2046,PhysRevE.50.888}, a pure state is in many ways 
indistinguishable from a thermal state. 
  
It is also worth mentioning that EOW branes may also be used in a dynamical manner, which means they appear as loops and are summed over in the path integral. In this approach they may provide an ingredient in defining a UV completion of JT gravity and solve the factorisation problem \cite{Gao:2021uro,Harlow:2018tqv}. \footnote{For 
other approaches to possible non-perturbative completions of JT gravity see 
\cite{Johnson:2019eik,Johnson:2020exp,Johnson:2022wsr}.}
 
Motivated by this,  we consider the computation of multi-boundary partition functions and matter 
correlation functions in the presence of an EOW brane. While we adopt the techniques developed 
in \cite{Saad:2019lba} and  \cite{Yang:2018gdb} respectively, the modified result we obtain due to 
the presence of the EOW brane is expected to represent the aforementioned quantities in a 
single-sided black hole geometry. 
 
Indeed  the main concern  of the present  paper is the computation of the late time behaviour of 
complexity. We define this as the geodesic length connecting the EOW brane and the asymptotic 
boundary.\footnote{In the Lorentzian picture this replaces the bridge-to-nowhere of \cite{Susskind:2014jwa}.}
More concretely, this is calculated in JT gravity as a {\it quenched expectation value}. The 
qualitative behaviour remains the same as in the case of a two-sided black hole, namely, the 
complexity grows linearly at late times up to a time $t \sim e^{S_0}$ and subsequently saturates to a 
constant value. The value of this constant which is of ${\cal O}(e^{S_0})$ depends crucially on the 
tension of the EOW brane. 
 
Although we adopt the non-perturbative definition\footnote{This is non-perturbative by virtue of 
an analytic continuation of the Euclidean path integral.} of complexity from \cite{Iliesiu:2021ari}, 
we refrain from rewriting it in terms of the correlators as in 
\eqref{eq:nonperturbativetimedependentlengthintroduction}. The reason is, although the quantity 
structurally looks similar to the aforementioned correlators, the limits on $\Delta$ appearing in the 
definition are counter-intuitive and do not agree with the standard geodesic approximation to the 
two-point function. 
 
Therefore we rather use a modified version of the replica trick in order to compute the quenched expectation value of the length of the geodesic.\footnote{Following 
\cite{Iliesiu:2021ari}, we only consider non
self-intersecting geodesics.} This avoids the aforementioned ambiguity. Moreover
using the definition of variance engendered by the modified replica approach, we observe 
time-independent results at late times both for the two-sided and the one-sided geometries.
This is in contrast with the result for the variance presented in \cite{Iliesiu:2021ari} where the complexity 
is defined in terms of a two-point function
\eqref{eq:nonperturbativetimedependentlengthintroduction}.

Our paper is organised as follows. We will start by introducing the theory of interest in section 
\ref{ref:JTgravityandwavefunctions}. By use of the quantisation procedure in the presence of a 
boundary brane \cite{Gao:2021uro}, we construct various wavefunctions needed in building up 
different partition functions and of course the path integral, which describes the volume of the 
black hole interior for our setting. In this section we also consider matrix elements in the geodesic 
length basis on the Hilbert space produced by the EOW brane. More specifically, we calculate the 
off-diagonal elements showing that while we are describing a pure state, they still obey the ETH. 
In sections \ref{sec:partitionfunction} and \ref{sec:correlationfunctions} we construct the multi-
boundary partition function and the quantum gravity matter correlation functions respectively. We 
put the pieces together in section \ref{sec:complexity}, where we compute the complexity using 
the definition mentioned above. Then we also consider the variance of this quantity. We conclude 
in section \ref{sec:conclusion-outlook} with a couple of interesting questions and comments on 
work in progress.


\section{Lorentzian JT gravity with EOW Branes and Wavefunctions}\label{ref:JTgravityandwavefunctions}
In this section we use the canonical quantisation procedure first introduced in \cite{Harlow:2018tqv}, to construct different wavefunction expressions for JT gravity in the 
presence of an EOW brane. After reviewing the quantisation procedure in presence of a boundary 
brane \cite{Gao:2021uro}, we generalise the construction to compute wavefunctions for different 
configurations of the EOW brane on the disk and then for the trumpet. These quantities are the 
essential building blocks in the calculation of correlation functions as well as complexity in our 
setup. 

\subsection{The classical solution}

JT gravity is a two-dimensional theory of gravity with the Lorentzian action 
\cite{Jackiw:1984je,Teitelboim:1983ux}
\begin{align}\label{eq:LorentzianJTaction}
    S_{\text{JT}}=\frac{S_0}{2 \pi}\left( \int \sqrt{-g}R + 2 \int \sqrt{|h|} K \right) + \int \sqrt{-g} \phi 
   \left(R+2\right) + 2 \int \sqrt{|h|} \phi \left(K-1\right)\,,
\end{align}
where the first term is the topological Gauss-Bonnet term and $S_0$ is the ground state entropy. 
In addition, we add the action of an EOW brane, which is of the form:
\begin{equation}\label{eq:EOWbraneaction}
    S_{\text{Brane}}=\mu \int_{\text{Brane}}ds\,, 
\end{equation}
with $\mu$ being the brane tension. In
two spacetime dimensions, the eq \eqref{eq:EOWbraneaction} boils down to the action 
of a particle with mass $\mu$. The overall action is given by
\begin{equation}\label{eq:overallaction}
    S= S_{\text{JT}} + S_{\text{Brane}}\,.
\end{equation}
The corresponding equations of motion are 
\begin{equation}
    R+2=0\,,\quad \nabla_{\mu}\nabla_{\nu}\phi - g_{\mu \nu} \nabla^2 \phi + g_{\mu \nu} \phi=0\,.
\end{equation}
At the asymptotic AdS boundary, the boundary conditions are set by fixing the induced metric 
and the dilaton value \cite{Maldacena:2016upp,Jensen:2016pah,Engelsoy:2016xyb}
\be\label{eq:LorentzianAsymptoticBoundaryConditions}
  ds^2|_{\partial M} =- \frac{dt^2}{\epsilon^2}\,,\;\;\;\;\;\;\;\;\;\;\;\;
    \phi|_{\partial M} = \frac{\phi_b}{\epsilon}\,,
\ee
where $\epsilon$ is a holographic renormalisation parameter and we are interested in the limit $
\epsilon \rightarrow 0$.
Additionally, at the EOW brane the following conditions are set \cite{Penington:2019kki}
\be\label{eq:EOWboundarycondtions}
    K = 0\,,\;\;\;\;\;\;\;\;\;\; \partial_{n}\phi = \mu \,.
\ee
Here $\partial_n$ denotes the derivative normal to the EOW brane. The latter condition is essential 
in ensuring dynamical gravity on the EOW brane.

\subsection{Quantisation in presence of a brane}

Let us denote the normalised geodesic distance between the AdS boundary and the EOW brane 
by $L$. The Hilbert space may be constructed in terms of $L_2$-normalisable functions of $L$.
\footnote{This is referred to as the ``L-basis'' in \cite{Harlow:2018tqv}. The choice of this basis 
avoids the subtlety of defining a ``time operator" whose dual Hamiltonian is bounded from below. 
Furthermore, this choice also allows for a full phase space $\mathbb{R}^2$ without any 
restrictions on the phase space coordinates.} As the system may be thought of as a particle in a 
Morse potential, the Hamiltonian amounts to \cite{Gao:2021uro}
\be \label{eq:Hamiltonian}
H=\frac{2}{\phi_b}\left(\frac{P^2}{4}+\mu e^{-L}+e^{-2L}\right)\,,
\ee
such that the Schr\"odinger equation is given by \cite{Harlow:2018tqv,Gao:2021uro}
\be\label{eq:ShE1}
\left(-\partial^2_L+4 \mu e^{-L}+4 e^{-2L}\right)\psi_{\mu,E}(L)=2 E\; \psi_{\mu,E}(L)\,.
\ee
In going from \eqref{eq:Hamiltonian} to \eqref{eq:ShE1} we have set $\phi_b=1$ and replaced $P
\rightarrow -i\partial_L$. In solving \eqref{eq:ShE1}, we are generally assuming $\mu >0$. Setting 
$k^2=2E$ and $z=4e^{-L}$ the corresponding normalised wavefunction 
\cite{Gao:2021uro} is \footnote{Due to the fact that $W_{a,b}=W_{a,-b}$ we are restricted to 
$k\geq 0$. } 
\be
\psi_{k,\mu}(z)=\sqrt{f_\mu(k)}\; \frac{W_{-\mu,ik}(z)}{\sqrt{z}},\;\;\;\;\;\;\;\; {\rm with}\;\;f_\mu(k)=
\gamma_\mu(k)r(k)\,,
\ee
where we have defined 
\be \label{eq:EOWfunction}
\gamma_\mu(k)=\left|\Gamma
\left(\frac{1}{2}+\mu+ik\right)\right|^2,\;\;\;\;\;\;\;\;\;\;r(k)=\frac{k\sinh(2\pi k)}{\pi^2}\,.
\ee
The normalisation of $\psi_{k,\mu}(z)$ requires the use of the orthogonality relation for Whittaker 
functions of the second 
kind of imaginary order \cite{Szmytkowski:2009}
\be\label{OW}
\int_0^\infty \frac{dz}{z^2} \;W_{-\mu,ik}(z)\; W_{-\mu, ik'}(z)=\frac{1}{f_\mu(k)}\delta(k-k')\,.
\ee
 The quantum mechanical propagator is \cite{Gao:2021uro}  
\bea \label{eq:quantumechanicalpropagator}
G_{\beta}(z_1,z_2)=\langle L_2 | e^{-\beta H} |L_1\rangle
=\int dk\, e^{-\frac{\beta k^2}{2}}\,f_\mu(k)\,\frac{W_{-\mu,ik}(z_1)}{\sqrt{z_1}} 
\frac{W_{-\mu, ik}(z_2)}{\sqrt{z_2}}\,.
\eea

Let us now come to a more geometric description in terms of the Euclidean path integral of 
JT gravity.  In the Euclidean picture, the time coordinate $\tau$ is periodic with 
$\tau \sim \tau + \beta$. The Euclidean action is given by 
\be\label{eq:EuclideanJTaction}
    S=-\frac{S_0}{2\pi}\left( \int \sqrt{g}R + 2 \int \sqrt{|h|} K \right) - \int \sqrt{g} \phi \left(R+2\right)
- 2 \int \sqrt{|h|} \phi \left(K-1\right)\,,
\ee
where we set the following boundary conditions for an asymptotic AdS boundary
\be\label{eq:EuclideanAsymptoticBoundaryConditions}
    ds^2|_{\partial M} = \frac{d\tau^2}{\epsilon^2}\,,\;\;\;\;\;\;\;\;\;\;\;
    \phi|_{\partial M} = \frac{\phi_b}{\epsilon}\,. 
\ee

Again, the first term of \eqref{eq:EuclideanJTaction} is purely topological and accounts for the 
Euler characteristic of the Riemann surface $\chi=2-2 g- n$, where $g$ is the genus and $n$ the 
number of boundaries. The integration over the dilaton localises the path integral to surfaces of 
constant negative curvature with an asymptotic boundary length determined by the boundary 
conditions \eqref{eq:EuclideanAsymptoticBoundaryConditions}. The extrinsic curvature term 
gives a Schwarzian action to the asymptotic boundary fluctuations on the hyperbolic space 
\cite{Maldacena:2016upp,Engelsoy:2016xyb}. 

The complete path integral includes an integral 
over the moduli of such surfaces and the boundary fluctuations. Briefly stated, the higher genus 
surfaces for one asymptotic boundary may be viewed as consisting of two parts, namely, one 
asymptotic boundary of fixed length and a geodesic boundary of length $b$ and a remaining 
genus $g$ Riemann surface with geodesic boundary of the same length $b$. The genus 
expansion of JT then takes on the form \cite{Saad:2019lba}:
\begin{equation}
    \langle Z(\beta) \rangle= e^{S_0} \hat{Z}_D(\beta) + \sum_{g=1} e^{(1-2g)S_0} 
    \int_{0}^{\infty} b db V_{g, 1}(b) \hat{Z}_T (\beta,b)\,,
\end{equation}
where $V_{g,1}$ is the Weil-Petersson volume of genus $g$ and one geodesic boundary 
\cite{Mirzakhani:2006fta,Eynard:2007fi} and the integration over $b$ glues the two parts of the 
surface together. Here $\hat{Z}_D(\beta)$ refers to the disk topology partition function and 
$\hat{Z}_T(\beta,b)$ to the ``trumpet" partition function \cite{Stanford:2017thb,Saad:2019lba}
\begin{equation}\label{eq:diskandtrumpetpartitionfunction}
    \hat{Z}_D(\beta)=\frac{e^{\frac{2\pi^2}{\beta}}}{\sqrt{ 2 \pi}\beta^{3/2}}\,,
    \quad\quad \hat{Z}_T(\beta,b)=\frac{e^{\frac{-b^2}{2 \beta}}}{\sqrt{ 2 \pi}\beta^{1/2}}\,.
\end{equation}
This construction can be generalised to $n$ asymptotic boundaries with the connected 
contribution being of the form \cite{Saad:2019lba}:
\begin{equation}\label{eq:standardJTpartitionfunctionmultipleboundaries}
    \langle Z(\beta_1)...Z(\beta_n)\rangle_{\text{C}}= \sum_{g=0} e^{(2- 2g-n)S_0}
    \hat{Z}_{g,n}(\beta_1,...,\beta_n)\,,
\end{equation}
with the definition
\begin{equation}
   \hat{Z}_{g,n}(\beta_1,...,\beta_n)= \int_{0}^{\infty}b_1 db_1...b_ndb_n V_{g,n}(b_1,...,b_n)
   \hat{Z}_T(\beta_1,b_1)... \hat{Z}_T(\beta_n,b_n)\,.
\end{equation}
Moreover, the hats, $\hat{}$ denote quantities without manifest topological weighting. 
Incorporating the latter, one defines $Z_D(\beta)=e^{S_0} \hat{Z}_D(\beta)$. 
In our construction, we additionally consider the addition of an EOW brane via the 
action \eqref{eq:EOWbraneaction} and the boundary conditions 
\eqref{eq:EOWboundarycondtions}. This modifies the partition function as we explain 
in the next sections.

At various points we will compute the expectation value of geodesic length in the Euclidean 
JT path integral. In contrast to the disk, on hyperbolic surfaces of genus $g \geq 1$ there are an 
infinite number of geodesics. Let us consider the case of non self-intersecting geodesics as in 
\cite{Iliesiu:2021ari}. The moduli space of hyperbolic, bordered Riemann surfaces 
$\mathcal{M}_{g,n}(b_1,...,b_n)$ comes with a symplectic form, the Weil-Petersson form 
$\Omega=\sum_{i=1}^{3g+n-3}db \wedge d\tau$, which in principle allows the calculation of the 
corresponding moduli space volume if restricted to a fundamental domain. Similarly, as first 
argued for in the $g=1$ case in \cite{Saad:2019pqd}, and elaborated upon in 
\cite{Blommaert:2020seb,Iliesiu:2021ari}, the integral of the geodesics over moduli space 
may be calculated by modding via the mapping class group, which we denote 
$\text{MCG}_{g,n}$. This leads to the expression 
\cite{Saad:2019pqd,Iliesiu:2021ari,Mirzakhani:2006fta,Iliesiu:2021ari}
\begin{equation}\label{eq:modulispaceformula}
\int_{\frac{\mathcal{M}_{g,1}}{\text{MCG}_{g,1}}} \Omega \sum_{\gamma} e^{-\Delta 
\ell_{\gamma}}=e^{-\Delta \ell}\int_{\frac{\mathcal{M}_{g-1,2}}{\text{MCG}_{g-1,2}}} \Omega  + 
\sum_{h \geq 0} e^{-\Delta \ell}\int_{\frac{\mathcal{M}_{h,1}}{\text{MCG}_{h,1}}} \Omega 
\int_{\frac{\mathcal{M}_{g-h,1}}{\text{MCG}_{g-h,1}}} \Omega\,.
\end{equation}
This formula may be visualised as cutting along the geodesic and considering the resulting 
geometries.

\subsection{The disk wavefunctions}

\begin{figure}[h]
     \centering
     \begin{subfigure}[b]{0.25\textwidth}
         \centering
         \includegraphics[width=\textwidth]{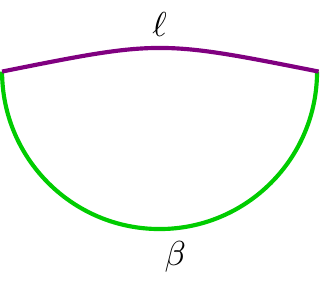}
         \caption{}
     \end{subfigure}
     \hfill
     \begin{subfigure}[b]{0.25\textwidth}
         \centering
         \includegraphics[width=\textwidth]{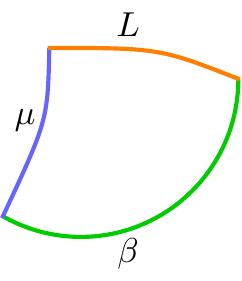}
         \caption{}
     \end{subfigure}
     \hfill
     \begin{subfigure}[b]{0.21\textwidth}
         \centering
         \includegraphics[width=\textwidth]{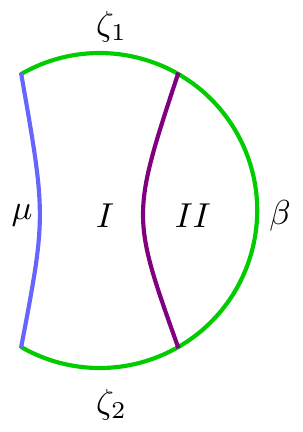}
         \caption{}
     \end{subfigure}
        \caption{Three possible disk configurations corresponding to different wavefunctions. 
Figure (a) is the wavefunction of the Hartle-Hawking state of the two-sided AdS system in JT 
gravity. Figure (b) and figure (c) are two options in the presence of an EOW brane. While in 
figure (b) we see a geodesic connecting the EOW brane to the asymptotic boundary, 
for figure (c) the geodesic connects to two different points on the asymptotic boundary. 
An orange curve corresponds to the former geodesic and a violet curve to the latter. 
Green denotes an AdS boundary and blue an EOW brane, respectively.}
\label{fig:diskwavefunction}
\end{figure}

Let us start by quickly revisiting some results we need from the two-sided AdS system. A natural 
procedure to prepare the states in the Hilbert space of the two-sided system is via the Hartle-
Hawking construction \cite{Harlow:2018tqv}, which is depicted in fig.\ref{fig:diskwavefunction}(a).

We denote the fixed geodesic length between two parts of the AdS boundary by $\ell$. 
Then the Hartle-Hawking wavefunction $\Phi_{D,\beta}(\ell)$ corresponds to the integral 
over all Euclidean geometries with disk topology and asymptotic AdS boundary of 
renormalised length $\beta$. Explicitly it amounts to
\be\label{eq:diskhartlehawkingwavefunction}
\Phi_{D}(\beta,\ell)
=2e^{S_0/2}\int_0^\infty dk e^{-\frac{\beta k^2}{2}}\,r(k)\; K_{2ik}(y)\,,
\ee
where $y=4 e^{-\frac{\ell}{2}}$. In this formalism the disk partition function is given as
\bea\label{eq:diskpartitionfunction}
&&Z_{D}(\beta)=\int_0^\infty \frac{dy}{y}\;\Phi_{D}(\beta/2,\ell)\;\Phi_{D}(\beta/2,\ell)=
\frac{e^{S_0}}{2}
\int_0^\infty dk\, e^{-\frac{\beta k^2}{2}}\,r(k)\cr &&\cr&&
\;\;\;\;\;\;\;\;\;=e^{S_0}\int_0^\infty dE\; e^{-\beta E}\hat{\rho}_{D}(E),
\eea
where $\hat{\rho}_{D}(E)$ is the disk density of states, which is given as 
\cite{Cotler:2016fpe,Bagrets:2017pwq,Stanford:2017thb,Mertens:2017mtv,Kitaev:2018wpr,
Yang:2018gdb}
\be\label{eq:diskdensityofstates}
\hat{\rho}_{D}(E)=\frac{\sinh(2\pi\sqrt{2E})}{2\pi^2}\,.
\ee
From \eqref{eq:diskpartitionfunction} we see that the wavefunction is normalised in such a way to 
give the correct expression for \eqref{eq:diskpartitionfunction} and \eqref{eq:diskdensityofstates}.

Before moving on to more complicated hyperbolic surfaces, let us now introduce the EOW 
brane already in this setting and construct the disk wavefunction in its presence. 
We can interpret the resulting wavefunction as the Hartle-Hawking wavefunction in the 
$L$ basis for the case of a one-sided black hole. 
This wavefunction is associated to a region enclosed by an asymptotically AdS boundary 
of renormalised length $\beta$, an EOW brane and a geodesic of length $L$ connecting them. 
\footnote{In contrast to the geodesic length connecting two points on the AdS boundary which we 
denoted by $\ell$.} This configuration is depicted in fig.\ref{fig:diskwavefunction}(b). 

We will denote the corresponding wavefunction by $\Psi_{D}(\beta,L)$ \footnote{We denote 
wavefunctions associated to the two-sided black hole via $\Phi$ and those in the presence of 
EOW branes by $\Psi$.} which should satisfy
\be\label{eq:eowwavefunctiondiskviapropagator}
\int_0^\infty \frac{dz}{z}\;\Psi_{D}(\beta/2,L)\Psi_{D}(\beta/2,L)=
\int_0^\infty \frac{dz_1}{z_1}\frac{dz_2}{z_2}\Psi_{D}(x,L_1)
G_{\beta-2x}(z_1,z_2)\Psi_{D}(x,L_2)\,,
\ee
where the variable of integration is $z = 4e^{-L}$.
It is straightforward to see that \eqref{eq:eowwavefunctiondiskviapropagator} is fulfilled for 
the following expression
\be\label{eq:eowwavefuntiondisk}
\Psi_{D}(\beta,L)
=\frac{e^{S_0/2}}{\sqrt{2}}\int_0^\infty dk \; e^{-\frac{\beta k^2}{2}}\,\gamma_\mu(k)r(k)\;
\frac{W_{-\mu,ik}(z)}{\sqrt{z}}\,.
\ee
The disk partition function in the presence of an EOW brane therefore amounts to
\begin{align}
\label{eq:diskpartitionfunctionwithEOWbrane}
Z_{D,\mu}(\beta) & = \int_0^\infty \frac{dz}{z}\;\Psi_{D}(\beta/2,L)\Psi_{D}(\beta/2,L)=
\frac{e^{S_0}}{2}\int_0^\infty dk
\;e^{-\frac{\beta k^2}{2}}\,\gamma_\mu(k)r(k)\nonumber \\
& = e^{S_0}\int_0^\infty dE\; e^{-\beta E}\gamma_\mu(E)\hat{\rho}_{D}(E),
\end{align}
Comparing \eqref{eq:diskpartitionfunctionwithEOWbrane} to \eqref{eq:diskpartitionfunction} 
we see that the effect of the EOW brane is encompassed by an additional $\Gamma$-function 
expression defined in \eqref{eq:EOWfunction}. The above expressions also allow us to calculate 
the wavefunction ${\Psi}_{{D}}(\zeta_1,\zeta_2,\ell)$ for region $I$ depicted in fig.
\ref{fig:diskwavefunction}(c): a region enclosed by an EOW brane and a geodesic connecting 
points on the asymptotic AdS boundary. This wavefunction can be derived from the identification,
\be
Z_{D,\mu}(\beta)=\int_0^\infty \frac{dy}{y}\;{\Psi}_{{D}}(\zeta_1,\zeta_2,\ell)
\Phi_{D}(\beta-\zeta_1-\zeta_2,\ell),
\ee
by which, using \eqref{eq:diskhartlehawkingwavefunction}, one arrives at
\be\label{eq:diskwavefunctionwithEOWandtwopointsasymptoticboundary}
{\Psi}_{{D}}(\zeta_1,\zeta_2,\ell)=2e^{S_0/2}\int_0^\infty dk\;
e^{-\frac{k^2}{2}(\zeta_1+\zeta_2)}\;\gamma_\mu(k)r(k)\;K_{2ik}(y)\,.
\ee


\subsection{The trumpet wavefunctions}
\label{sec:trumpetwf}
The most important ingredients of our study are the wavefunctions on the trumpets whose 
asymptotic boundaries are either pinched off by the disk regions considered in 
fig.\ref{fig:diskwavefunction} or replaced in some parts by the EOW brane.

\begin{figure}[h]
     \centering
     \begin{subfigure}[b]{0.23\textwidth}
         \centering
         \includegraphics[width=\textwidth]{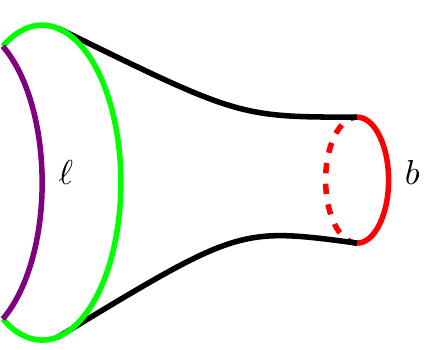}
         \caption{}
         \label{fig:y equals x}
     \end{subfigure}
     \hfill
     \begin{subfigure}[b]{0.25\textwidth}
         \centering
         \includegraphics[width=\textwidth]{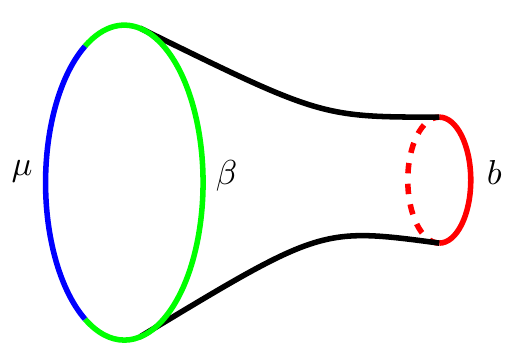}
         \caption{}
     \end{subfigure}
     \hfill
     \begin{subfigure}[b]{0.23\textwidth}
         \centering
         \includegraphics[width=\textwidth]{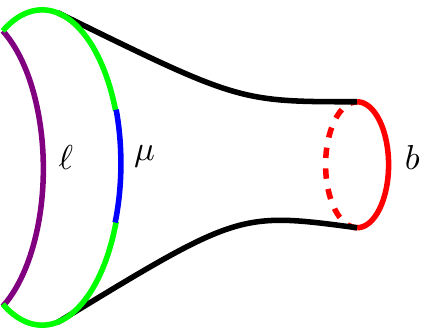}
         \caption{}
     \end{subfigure}
     \hfill
     \begin{subfigure}[b]{0.25\textwidth}
         \centering
         \includegraphics[width=\textwidth]{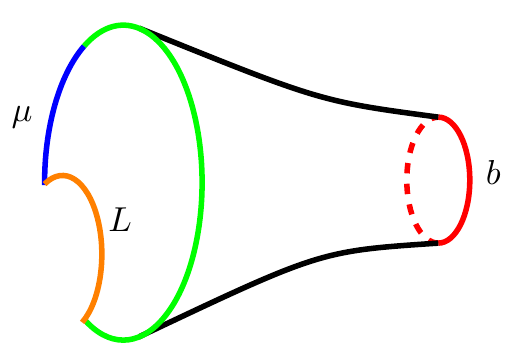}
         \caption{}
     \end{subfigure}
        \caption{Four different possible trumpet geometries corresponding to four distinct 
wavefunctions. The closed geodesic boundary is depicted in red. In figure (a) we see the 
generalisation of the disk configuration, figure \ref{fig:diskwavefunction}(a) to the trumpet. Figure 
(b) corresponds to the wavefunction on a trumpet geometry with both an EOW brane and an 
asymptotically AdS boundary. Figure (c) shows the wavefunction of a geodesic connecting two 
points on an asymptotically AdS boundary which contains an EOW brane. Lastly, in figure (d) we 
see a geodesic connecting EOW brane and AdS boundary on a trumpet geometry.}
        \label{fig:trumpetwavefunctions}
\end{figure}

While more complicated hyperbolic surfaces require the use of Riemann surfaces with geodesic 
boundaries, the simplest configuration on the trumpet is depicted in fig.
\ref{fig:trumpetwavefunctions}(a).

The corresponding wavefunction $\Phi_T(\beta,b,\ell)$ can be realised as the trumpet 
wavefunction pinched off by the disk wavefunction shown in fig. \ref{fig:diskwavefunction}(a). 
This is obtained through the identity
\be 
\Phi_{T}(\beta,b)=\frac{1}{\pi}\int_0^\infty dk\, \cos(kb)\,e^{-\frac{\beta k^2}{2}}=
\int_{0}^{\infty} \frac{dy}{y}\;\Phi_T(\beta-x,b,\ell)\; \Phi_{D}(x,\ell)\,,
\ee
which results in
\be\label{eq:trumpetwavefunction}
\Phi_T(\beta,b,\ell)=\frac{4e^{-S_0/2}}{\pi}\int_0^\infty dk\;e^{-\frac{\beta k^2}{2}}\,\cos(kb)\; 
K_{2ik}(y)\,.
\ee

Let us now come to the geometry depicted in fig.\ref{fig:trumpetwavefunctions}(b). This can be 
computed by gluing the above geometry with a region enclosed by a geodesic and EOW brane 
as shown in fig.\ref{fig:diskwavefunction}(c). This yields the wavefunction $\Psi_{T}(\beta,b)$ 
associated with this diagram
\be\label{eq:trumpetwavefunctionwitheow}
\Psi_{T}(\beta,b)=\int_0^\infty \frac{dy}{y}\;\Phi_T(\beta,b,\ell)\;{\Psi}_{{D}}(0,0,\ell)=
\frac{1}{\pi}\int_0^\infty dk \; 
\cos(kb)\; \gamma_\mu(k)\,e^{-\frac{\beta k^2}{2}}.
\ee
The wavefunction \eqref{eq:trumpetwavefunctionwitheow} is in a perfect agreement with the 
corresponding wavefunction presented in \cite{Gao:2021uro}. 

This in turn allows for the calculation of $\Psi_{T}(\zeta_1,\zeta_2,b,\ell)$, the wavefunction 
associated with fig.\ref{fig:trumpetwavefunctions}(c) and obtained through the equation
\be 
\Psi_{T}(\beta,b)=\int_0^\infty \frac{dy}{y}\;\Psi_{T}(\zeta_1,\zeta_2,b,\ell)
\Phi_{D}(\beta-\zeta_1-\zeta_2,\ell)\,.
\ee
Using \eqref{eq:diskhartlehawkingwavefunction} and \eqref{eq:trumpetwavefunctionwitheow}, this yields
\be\label{eq:trumpeteowwithtwoboundarypoints}
\Psi_{T}(\zeta_1,\zeta_2,b,\ell)=\frac{4e^{-S_0/2}}{\pi}\int_0^\infty dk \; 
\cos(kb)\,\gamma_\mu(k)\,e^{-\frac{k^2}{2}(\zeta_1+\zeta_2)}\,K_{2ik}(y)\,.
\ee
Finally, the wavefunction corresponding to the geometry shown in the  panel (d) of fig.
\ref{fig:trumpetwavefunctions}, namely a trumpet geometry with geodesic of length $L$ from the 
EOW brane to the asymptotic boundary, can be computed by pinching-off the wavefunction $
\Psi_{D}(\beta,L)$ from the
above wavefunction. Therefore the structure
\be\label{eq:trumpetwavefunctionwitheowandgeodesicabstract}
\Psi_{T}(\beta,b)=\int_0^\infty \frac{dz}{z}\;\Psi_{T}(\beta-x,b,L)\;\Psi_{D}(x,L)\,,
\ee
by use of \eqref{eq:eowwavefuntiondisk} results in the wavefunction
\be\label{eq:trumpetwavefunctionwitheowandgeodesic}
\Psi_{T}(\beta,b,L)=\frac{\sqrt{2}e^{-S_0/2}}{{\pi}}\int_0^\infty dk\,
\cos(k b)\gamma_\mu(k)\,
e^{-\frac{\beta k^2}{2}}\,z^{-1/2}\,W_{-\mu,ik}(z)\,.
\ee

\subsection{Pure vs. Thermal States}

As already mentioned in the introduction, by considering an EOW brane we are describing a pure 
state. However, to establish its interpretation as a typical boundary state, it is essential to try and 
delineate differences to a thermal state. We can check the expectation value of the energy. 
Indeed at disk level this amounts to 
\be\label{eq:energyexpectationvalue}
\langle E \rangle =\frac{\int_0^\infty \frac{dz}{z}\,\Psi_D(\beta/2,L)\, H\, \Psi_D(\beta/2,L)}{
\int_0^\infty \frac{dz}{z}\,\Psi_D(\beta/2,L)\,\Psi_D(\beta/2,L)},
\ee
with $H$ being the Hamiltonian defined in \eqref{eq:Hamiltonian}. As the corresponding system 
may be thought of as a particle in a Morse potential, by use of the Schr\"odinger equation, one 
arrives at
\be\label{eq:energyexpectationvaluepart2}
\langle E \rangle =-\frac{\partial}{\partial\beta}\ln  Z_\mu(\beta)\,,
\ee
which is in agreement with the expectation value of a thermal ensemble with temperature 
$\frac{1}{\beta}$. This may be readily generalised to higher genus. Therefore the wavefunctions 
in the presence of an EOW brane indeed correspond to states which are indistinguishable from 
thermal states. 

On the other hand we note that the ETH delineates between diagonal and off-diagonal matrix 
elements. More explicitly, the matrix elements of observables in the eigenstate of the Hamiltonian 
are given by \cite{1999}
\begin{equation}\label{eq:ETH}
  \mathcal{O}_{m n}=  \mathcal{O}(\overline{E}) \delta_{m n} + e^{-\frac{S(\overline{E})}{2}} 
  f_{ \mathcal{O}}\left(\overline{E},\omega\right) R_{m n}\,,  
\end{equation}
where $\overline{E}=\frac{E_m+E_n}{2}$, $\omega=E_m - E_n$ and $S(\overline{E})$ 
is the entropy. Moreover, $\mathcal{O}(\overline{E})$ is the expectation value in the 
microcanonical ensemble, $f_{ \mathcal{O}}\left(\overline{E},\omega\right)$ is a smooth function 
and $R_{m n}$ a random variable with zero mean and unit variance. 

\begin{figure}
\begin{center}
\includegraphics[scale=0.8]{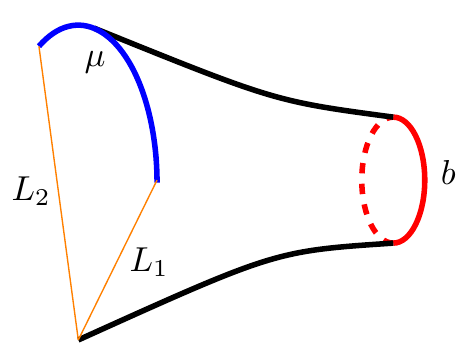}\;\;\;\;\;\;\;\;\;\;\;
\end{center}
\caption{One important ingredient in the calculation of the leading order correction to the inner product $\langle L_1| L_2 \rangle$ via Euclidean path integral. We see the two geodesics $L_1$ and $L_2$ in orange, the EOW brane in blue and a closed geodesic $b$ in red. The wavefunction of this geometry is denoted by $\Psi_{T}(b,L_1,L_2)$. Topologies beyond the disk are important in recovering ETH-like behaviour.}
\label{fig:innerproductwavefunction}
\end{figure}

One observes  that off-diagonal 
elements are suppressed by the Hilbert space size. In order to show that the wavefunction we 
consider also satisfies ETH, we need to calculate off-diagonal elements of the inner product in the 
length basis $|L\rangle$ used in the quantisation of \eqref{eq:Hamiltonian}. Actually the inner 
product we need for this analysis was already considered in \cite{Gao:2021uro}, where the 
importance of higher topologies was stressed. First we need to define a building block, which is 
shown in fig.\ref{fig:innerproductwavefunction}.
Denoting the corresponding wavefunction by $\Psi_{T}(b,L_1,L_2)$, one has
\be
\Psi_{T}(\beta,b)=\int_0^\infty\frac{dz_1}{z_1}\;\frac{dz_2}{z_2} \Psi_{D}(\beta-x,L_1)
\Psi_{T}(b,L_1,L_2)\Psi_{D}(x,L_2)\,,
\ee
which in combination with the expression \eqref{eq:eowwavefuntiondisk} may be used to find
\be \label{eq:wavefunctionforinnerproduct}
\Psi_{T}(b,L_1,L_2)=\frac{2e^{-S_0}}{\pi}\int_0^\infty dk\;\cos(kb)\,\gamma_\mu(k)
\;(z_1z_2)^{-1/2}W_{-\mu,ik}(z_1)\;W_{-\mu,ik}(z_2)\,,
\ee
in agreement with the result obtained in \cite{Gao:2021uro}. 
The wavefunction \eqref{eq:wavefunctionforinnerproduct} plays an important role in recovering 
ETH behaviour, as the standard canonical quantisation condition
\begin{equation}
    \langle L_1 | L_2 \rangle= \delta \left(L_1-L_2\right)\,,
\end{equation}
is corrected via higher genus contributions to the expression
\be\label{eq:highergenuscontributiontoquantisationconditionpart1}
\langle L_1|L_2\rangle=\delta(L_1-L_2)+\int_0^\infty b\,db\,X(b)\,\Psi_T(b,L_1,L_2),
\ee
where we have introduced the notation $X(b)$ as in \cite{Gao:2021uro}. 
Here $X(b)$ is an integration measure which corresponds to all topologies ending 
on a single closed geodesic length $b$, such that the weighting by the Euler 
characteristic and the Weil-Petersson volumes are included in this quantity. 
We could also consider it to include an arbitrary number of EOW brane loops as in 
\cite{Gao:2021uro}. By use of \eqref{eq:wavefunctionforinnerproduct}, 
\eqref{eq:highergenuscontributiontoquantisationconditionpart1} takes on the form
\be \label{eq:eq:highergenuscontributiontoquantisationconditionpart2}
\langle L_1|L_2\rangle=\delta(L_1-L_2)+\frac{2e^{-S_0}}{\pi}\int_0^\infty dk\;
\chi(k)\,\gamma_\mu(k)
\;\frac{W_{-\mu,ik}(z_1)\;W_{-\mu,ik}(z_2)}{\sqrt{z_1z_2}}\,,
\ee
where
\be
\chi(k)=\int_0^\infty b\,db\, X(b)\,\cos(kb)\,.
\ee
The leading contribution to the off-diagonal term comes from surfaces with genus one for which $\chi(k)\sim e^{-S_0}$, which results in
\be
\langle L_1|L_2\rangle\approx \delta(L_1-L_2)+(\cdots)_{L_1,L_2} \,e^{-2S_0}\,,
\ee
in agreement with \cite{Miyaji:2021ktr}. Here $(\cdots)_{L_1,L_2}$ refers to the $g=1$ contribution, 
where we have already pulled out the topological weighting. We therefore see that off-diagonal 
terms are suppressed exponentially just as in \eqref{eq:ETH}.


\section{Partition Function}\label{sec:partitionfunction}

\begin{figure}
\begin{center}
\includegraphics[scale=0.8]{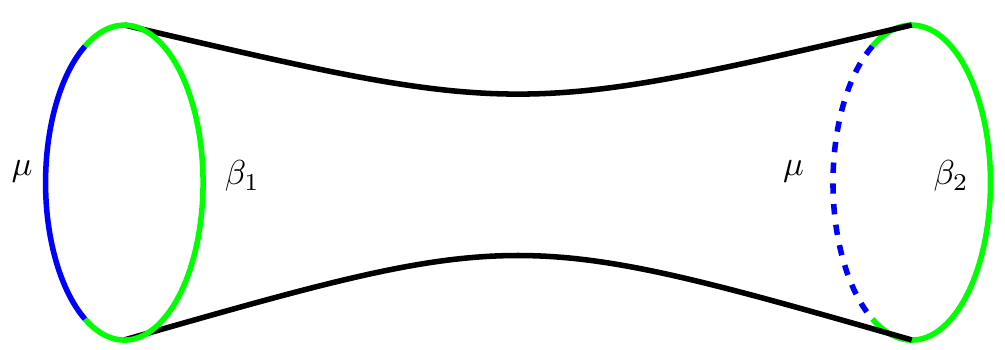}\;\;\;\;\;\;\;\;\;\;\;
\end{center}
\caption{Two trumpet geometries glued  together along their closed geodesic boundaries. This 
geometry corresponds to the connected part of the spectral form factor.  } 
\label{fig:2trumpet}
\end{figure}

In this section we construct the partition function in the presence of an EOW brane via 
the wavefunction formalism developed in section \ref{ref:JTgravityandwavefunctions}. 
The most natural quantity to analyse is the two-point function or the spectral form factor.
 More specifically, we require the trumpet wavefunction \eqref{eq:trumpetwavefunctionwitheow}. 
 We may visualise the connected contribution to the two-point function as gluing two 
 trumpet geometries of the type illustrated in fig.\ref{fig:trumpetwavefunctions}(a) together 
 along their closed geodesic boundaries, which results in the geometry shown in 
 fig.\ref{fig:2trumpet}.

In analogy to \eqref{eq:standardJTpartitionfunctionmultipleboundaries}, the overall contribution 
including connected and disconnected structures gives the following expression:
\begin{align}\label{eq:JTtwopointfunction}
&\langle Z(\beta_1)Z(\beta_2)\rangle_{\mu}\nonumber\\=& \int_0^\infty b_1db_1\;b_2db_2\;
\Psi_T(\beta_1,b_1)X(b_1,b_2)\Psi_T(\beta_2,b_2)\\
=&\frac{e^{-S_0}}{\pi^2}\int_0^\infty dk_1dk_2\;
e^{-\frac{\beta_1 k_1^2}{2}-\frac{\beta_2 k_2^2}{2}}
\gamma_\mu(k_1)\gamma_\mu(k_2)
\int_0^\infty b_1db_1 b_2db_2 X(b_1,b_2) 
{\cos(k_1b_1)\cos(k_2b_2)}.\nonumber
\end{align}
Here we have introduced the function  $X(b_1,b_2)$ that denotes the topologically weighted sum 
over the Weil-Petersson volumes associated to surfaces with two geodesic boundaries 
parametrised by $b_1$ and $b_2$. It is of the form
\be\label{eq:X}
X(b_1,b_2):= \sum_{g=0}e^{(2-2g) S_0} \left({ V}_{g-1,2}(b_1,b_2)+\sum_{a\geq 0}
{ V}_{g-a,1}(b_1)
{V}_{a,1}(b_2)\right)\,.
\ee
We note that the first term of \eqref{eq:X} corresponds to the connected contribution, whereas the 
second term corresponds to the disconnected contribution. There are two contributions in 
\eqref{eq:JTtwopointfunction} which must be put in ``by hand" as the moduli space volumes 
$V_{g=0,1}(b)$ and $V_{g=0,2}(b_1,b_2)$ in \eqref{eq:X} are undefined.\footnote{These two 
volumes constitute input values for the topological recursion 
\cite{Eynard:2004mh,Eynard:2007kz}.} For the disconnected contributions involving 
$V_{g=0,1}(b)$, the correct result is given by \eqref{eq:diskpartitionfunctionwithEOWbrane} 
ands the two boundary $g=0$ connected contribution is defined as
\begin{align}
     Z(\beta_1,\beta_2)_{g=0,n=2,\mu} &=\int_{0}^{\infty}b_1 db_1 b_2 db_2 
     \Psi_T(\beta_1,b_1) \Psi_T(\beta_2,b_2)\,.
\end{align}
Comparing \eqref{eq:JTtwopointfunction} to the two-sided expression of \cite{Saad:2019lba}, 
one observes   that the distinction to \eqref{eq:JTtwopointfunction} lies in the factor 
$\gamma_\mu(k_1)\gamma_\mu(k_2)$. Analytically continuing
 \eqref{eq:JTtwopointfunction} to the spectral form factor and rewriting in terms of 
 energy variables one arrives at
\be\label{eq:SFF}
\langle Z(\beta+it)Z(\beta-it)\rangle_{\mu}=\int_0^\infty dE_1dE_2
e^{-\beta( E_1+E_2)-it(E_1- E_2)}\;
\gamma_\mu(E_1)\gamma_\mu(E_2)\;
\langle \rho(E_1) \rho(E_2) \rangle
\ee
where 
\be\label{eq:twopointfunctiondensity}
\langle \rho(E_1) \rho(E_2) \rangle = \int b_1db_1\,b_2db_2\,
X(b_1,b_2) \; 
\frac{\cos(b_1\sqrt{2E_1})\cos(b_2\sqrt{2E_2})}{2\pi^2\sqrt{E_1E_2}},
\ee
which is the density of states corresponding to two boundary case of 
\eqref{eq:standardJTpartitionfunctionmultipleboundaries}. At late times, the integral 
\eqref{eq:SFF} is dominated by small energy ranges, and it can be shown that for  $|E_1-E_2|\ll1$, 
non-perturbative contributions give the following expression for \eqref{eq:twopointfunctiondensity}\cite{Saad:2019lba}\footnote{See also 
\cite{Altland:2020ccq} based on the elegant approach of \cite{Wegner,Efetov:1983xg}.}
\be \label{eq:sinekernel}
\langle \rho(E_1) \rho(E_2) \rangle \approx e^{2 S_0}\hat{\rho}_D(E_1)
\hat{\rho}_D(E_2)+e^{S_0}\hat{\rho}_D(E_2)\delta(E_1-E_2)
-\frac{\sin^2\left(\pi e^{S_0} \hat{\rho}_D(E_2)(E_1-E_2)\right)}{\pi^2 (E_1-E_2)^2}\,,
\ee
where ${\hat \rho}_D(E)$ refers to the genus zero contribution to the density of states 
\eqref{eq:diskdensityofstates}. The last term in \eqref{eq:sinekernel} is the so-called sine-kernel. 
The non-perturbative nature of this contribution can be spotted by noting the factor of 
$e^{S_0}$ inside the ``$\sin$". As should be expected, plugging \eqref{eq:sinekernel} into 
\eqref{eq:SFF}, gives a ramp-plateau structure for the connected and decaying behaviour for the 
disconnected contribution.


\section{Correlation Functions}\label{sec:correlationfunctions}

\begin{figure}
\begin{center}
\includegraphics[scale=1.]{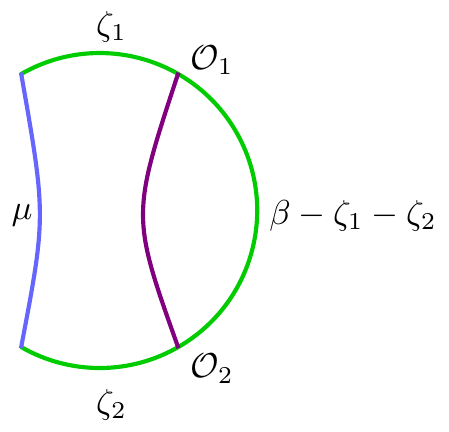}\;\;\;\;\;\;\;\;\;\;\;
\end{center}
\caption{This figure corresponds to the two-point function of two operators $\mathcal{O}_1$ and $
\mathcal{O}_2$ on the Euclidean disk in the presence of an EOW brane. The corresponding 
wavefunction is given in \eqref{eq:diskwavefunctionwithEOWandtwopointsasymptoticboundary}.} 
\label{fig:2pointfunction}
\end{figure}

Following the procedure of \cite{Yang:2018gdb} we will now determine full quantum gravity 
expressions for the matter correlation functions in the presence of an EOW brane. The idea of 
\cite{Yang:2018gdb} is to construct a certain Kernel which can be used to dress quantum field 
theory correlation functions on AdS$_2$ to produce gravity correlators. For the two-sided case, 
the Kernel essentially amounts to the Hartle-Hawking wavefunction 
\eqref{eq:diskhartlehawkingwavefunction}. More concretely, let us denote the coordinates by 
${\bf x}=(\xi,x)$, where $\xi$ is the holographic coordinate and $x$ the boundary coordinate. 
The regularised geodesic distance between two points is given by
\be\label{eq:regularisedgeodesicdistance}
e^{\frac{\ell}{2}}=\frac{|x_1-x_2|}{\sqrt{\xi_1\xi_2}}.
\ee
In terms of this expression the Kernel is
\be\label{eq:HartleHawkingKernel}
K(u_{12},{\bf x}_1,{\bf x}_2) = 2e^{S_0/2} \frac{4\sqrt{\xi_1\xi_2}}{|x_1-x_2|} 
\int_0^\infty dk e^{-\frac{u_{12} k^2}{2}}\,r(k)\; 
K_{2ik}\left( \frac{4\sqrt{\xi_1\xi_2}}{|x_1-x_2|}  \right)\,.
\ee
The quantum gravity correlators constructed in \cite{Yang:2018gdb} then amount to
\bea\label{eq:quantumgravitycorrelators}
\langle \mathcal{O}_1(u_1)\cdots \mathcal{O}_n(u_n)\rangle_D= \int_{x_1>\cdots>x_n}
\frac{\prod_{i}d\xi_idx_i}{{\rm Vol}\left(\text{SL}(2,R)\right)}\!&&\!K(u_{12},{\bf x}_1,{\bf x}_2)\cdots
K(u_{1n},{\bf x}_n,{\bf x}_1)\\ &&\!\times\,\prod_i \xi_i^{\Delta_i-2}\langle \mathcal{O}_1(x_1)\cdots 
\mathcal{O}_n(x_n)\rangle_{\rm CFT}\,,\nonumber
\eea
where $\Delta_i$ is the scaling dimension of the operator $\mathcal{O}_i$. 
${{\rm Vol}\left(\text{SL}(2,R)\right)}$ reminds us that one needs to fix the $\text{SL}(2, R)$ 
gauge symmetry. In our case, while the general logic leading to the structure of 
\eqref{eq:quantumgravitycorrelators} is preserved, now two different Kernels have to be 
used. In addition to \eqref{eq:HartleHawkingKernel}, a Kernel must be introduced due to the 
presence of the EOW brane. A quick look at fig.\ref{fig:2pointfunction} suggests that this Kernel 
corresponds to the wavefunction 
\eqref{eq:diskwavefunctionwithEOWandtwopointsasymptoticboundary}, which results in the 
expression
\be \label{eq:EOWbraneKernel}
M(\zeta_1,\zeta_2,{\bf x}_1, {\bf x}_2)=2e^{S_0/2}
\frac{4\sqrt{\xi_1\xi_2}}{|x_1-x_2|} 
\int_0^\infty\! dk\;
e^{-\frac{k^2}{2}(\zeta_1+\zeta_2)}\;\gamma_\mu(k)r(k)\;K_{2ik}\left( \frac{4\sqrt{\xi_1\xi_2}}
{|x_1-x_2|}  \right)\,.
\ee
Using this kernel and \eqref{eq:HartleHawkingKernel} the quantum gravity correlators in the 
presence of an EOW brane is
\begin{align}
\label{eq:quantumgravitcorrelatorswithEOW}
\langle \mathcal{O}_1(u_1)\cdots \mathcal{O}_n(u_n)\rangle_{D,\mu} &=\int_{x_1>\cdots>x_n}
\frac{\prod_{i}d\xi_idx_i}{{\rm Vol}\left(\text{SL}(2, R)\right)}K(u_{12},{\bf x}_1,{\bf x}_2)\cdots
K(u_{n-1\,n},{\bf x}_{n-1},{\bf x}_n) \nonumber\\ 
&\times M(\zeta_1,\zeta_n,{\bf x}_n, {\bf x}_1)\prod_i \xi_i^{\Delta_i-2}\langle 
\mathcal{O}_1(x_1)\cdots \mathcal{O}_n(x_n)\rangle_{\rm CFT}.
\end{align}

The above expressions are for disk topology as indicated by the index $D$. Let us briefly 
describe how to generalise to arbitrary topology by use of the two-point function as a concrete 
example. For the disk the two-point function is shown in fig.\ref{fig:2pointfunction}. The 
variables of fig.\ref{fig:2pointfunction} are related to those of formula 
\eqref{eq:quantumgravitcorrelatorswithEOW} via $u=\zeta_1 + \zeta_2$. Keeping in mind that the 
CFT two-point function is given by:
\be
\langle \mathcal{O}_1(x_1)\mathcal{O}_2(x_2)\rangle= e^{-\Delta \ell}\,,
\ee
we arrive at the quantum gravity two-point function at disk level (according to \eqref{eq:quantumgravitcorrelatorswithEOW})
\bea\label{eq:quantumgravitydisk2pf}
\langle \mathcal{O}_1(\zeta_1+\zeta_2)\mathcal{O}_2(0)\rangle_{D,\mu}\!&\!=\!&\!
\int_0^\infty \frac{dy}{y}\;\Phi_{D}(\beta-\zeta_1-\zeta_2,\ell)\Psi_{D}(\zeta_1,\zeta_2,\ell)\;
\left(\frac{y}{4}\right)^{2\Delta}
\\
\!&\!=\!&\!e^{S_0}\int_0^\infty dk_1\, dk_2\; 
e^{-\frac{k_1^2}{2}(\beta-u)-\frac{k_2^2}{2}u}\;
r(k_1)r(k_2)\,\gamma_\mu(k_2)\; {\cal N}(\Delta, k_1,k_2)\,,\nonumber
\eea
where 
\be\label{eq:Besselintegrationfunction}
{\cal N}(\Delta, k_1,k_2)=4\int_0^\infty \frac{dy}{y}\;
K_{2ik_1}\left( y \right)
K_{2ik_2}\left( y \right) \left(\frac{y}{4}\right)^{2\Delta}=\frac{|\Gamma(\Delta+i(k_1+k_2))
\Gamma(\Delta+i(k_1-k_2))|^2}{2^{2\Delta+1}\Gamma(2\Delta)}\,.
\ee
Comparing \eqref{eq:quantumgravitydisk2pf} to the expression for the two-sided AdS black hole 
\cite{Yang:2018gdb} we again see the new factor $\gamma_{\mu}\left(k\right)$ due to the 
presence of the EOW brane. 

In order to generalise \eqref{eq:quantumgravitydisk2pf} to higher genus, the wavefunctions 
\eqref{eq:trumpetwavefunction} and \eqref{eq:trumpeteowwithtwoboundarypoints} are needed. 
By making use of these wavefunctions and formula \eqref{eq:modulispaceformula} the two-point 
function of arbitrary genus is
\bea\label{eq:quantumgravity2pfirstexpression}
\langle \mathcal{O}_1(\zeta_1+\zeta_2)\mathcal{O}_2(0)\rangle_{\mu}&=&\int b_1db_1\,b_2db_2\,
X(b_1,b_2)\\ &&\;\;\;\;\;\;\;\;\;\;\;\times
\int_0^\infty \frac{dy}{y}\;\Phi_{T}(\beta-\zeta_1-\zeta_2,b_1,\ell)\Psi_{T}
(\zeta_1,\zeta_2,b_2,\ell)\;
\left(\frac{y}{4}\right)^{2\Delta}\nonumber\\
&=&\frac{16 e^{- S_0}}{\pi^2}\int_0^\infty dk_1 dk_2\;
\gamma_\mu(k_2)  e^{{-\frac{k_1^2}{2}(\beta-u)}{-\frac{k_2^2}{2}u}} {\cal N}(\Delta,k_1,k_2)
\nonumber \\  &&\;\;\;\;\;\;\;\;\;\;\;\;\;
\times\;\int b_1db_1\,b_2db_2\,
X(b_1,b_2) \; 
{\cos(k_2b_2)\cos(k_1b_1)}.\nonumber
\eea
We note, however, that  the disk contribution is a particular case and it is understood that the genus 
zero contribution is defined to be \eqref{eq:quantumgravitydisk2pf}. Altogether one gets 
 \be
\langle \mathcal{O}_1(u)\mathcal{O}_2(0)\rangle_{\mu}=16e^{-S_0}\int_0^\infty dE_1 dE_2
 e^{-E_1(\beta-u)-E_2 u}\; \gamma_\mu(E_2)\; \langle \rho(E_1) \rho(E_2) \rangle\; {\cal N}
 (\Delta,E_1,E_2) \,,
\ee 
where we are using \eqref{eq:twopointfunctiondensity}. The late-time behaviour of the two-point 
function amounts to considering the analytic continuation $u=\beta + i t$, which gives
\be
\langle \mathcal{O}_1(t)\mathcal{O}_2(0)\rangle_{\mu}=16e^{-S_0}\int_0^\infty dE_1 dE_2
 e^{-\frac{\beta}{2}(E_1+E_2)+it(E_1-E_2)}\; \gamma_\mu(E_2)\; \langle \rho(E_1) \rho(E_2) \rangle
 \; {\cal N}(\Delta,E_1,E_2) \,.
\ee
Comparing this expression to \eqref{eq:SFF} shows that the late-time behaviour is essentially the 
same as that of spectral
form factor. Indeed as far as the ramp and the plateau are concerned the extra
 ${\cal N}(\Delta,E_1,E_2) $ plays no essential role.


\section{The late time behaviour of complexity}\label{sec:complexity}

In this section we would like to study the late time behaviour of complexity in our setup. 
It is conjectured that the holographic quantum complexity is given by the volume of the Einstein-Rosen bridge \cite{Stanford:2014jda}. In our language in two dimensions it translates into  the length of a 
geodesic connecting two boundaries. This definition was used to compute the late time 
behaviour of complexity of a two-sided 
black hole in \cite{Iliesiu:2021ari}. In that work it was shown that the complexity 
exhibits linear growth at late times before it eventually saturates to a finite value. As detailed in the 
introduction, the most essential step in this construction was the use of the non-perturbative expression \eqref{eq:sinekernel} to furnish the saturation at late times.

In this section we adopt the same logic to work out the late time behaviour of complexity 
for a single-sided black hole. Crucially however, we do not relate the geodesic length to a matter two-point function but use the quenched expectation value. For the calculation of the complexity itself this leads to the same expressions for the two-sided case but a decisively different result for the variance. For the one-sided case, we need to 
compute the quenched expectation value of a geodesic suspended between the AdS boundary and the EOW brane.
Note that, in our notation, classically the geodesic distance 
between boundary and EOW brane is denoted by $L=- \ln z/4$. 
The complexity is therefore proportional to the expectation value of 
the geodesic ${\cal C}\sim \langle L\rangle_{\rm QG}$ in quantum gravity.
It is also worth noting that in the present case one could also compute the expectation 
value of a geodesic length connecting two points on the boundary, $\langle \ell\rangle_{QG}$. In what follows we will study the time dependence of these 
quantities using the
wavefunction formalism we developed in the previous sections.

\subsection{The geodesic $\ell$}\label{sec:labell}
To proceed, let us start with the  
geodesic $\ell$ which is used in the two-sided case and compute its ``quantum expectation'' 
value.  At the disk level one has
\be
\langle \ell(u)\rangle =-\frac{1}{Z_{D,\mu}(\beta)}\int_0^\infty \frac{dy}{y}\;\Psi_{D}
(\zeta_1,\zeta_2,\ell) \;\Phi_{D}
(\beta-u,\ell)\;
(2\ln \frac{y}{4}),\;\;\;\;\;\;\;\;{\rm with}\;u=\zeta_1+\zeta_2\,.
\ee 
To evaluate this quantity, we will use a trick which is inspired by the replica trick used {\it e.g.} 
in computing the quenched free energy. We write the logarithm in terms of the following limit
\footnote{In the context of JT gravity, see \cite{Engelhardt:2020qpv, Johnson:2020mwi, Johnson:2021rsh, Alishahiha:2020jko}.} 
\be\label{RT}
\ln A=\lim_{N\rightarrow 0}\frac{A^N-1}{N}= \lim_{N\rightarrow 0}\frac{d}{dN}\;A^N\,.
\ee

We normalise  by multiplying with a factor of ${Z_{D,\mu}^{-1}(\beta)}$, where $Z_{D,\mu}(\beta)$ is the disk partition 
function, given in 
\eqref{eq:diskpartitionfunctionwithEOWbrane}. Using this definition one may define complexity as
\be
\label{eq:replica-formula}
\langle \ell(u)\rangle =-\lim_{N\rightarrow 0}\frac{\langle y^{2N}\rangle_u-1}{N},
\ee
where
\be\label{yN}
\langle y^{2N}\rangle_u=\frac{1}{Z_{D,\mu}(\beta)}\int_0^\infty \frac{dy}{y}\;\Psi_{D}
(\zeta_1,\zeta_2,\ell) \;\Phi_{D} (\beta-u,\ell)\;\left( \frac{y}{4}\right)^{2N}
\ee
and it is understood that an analytic continuation must still be performed.
Expressions such as \eqref{yN} may then be calculated via \eqref{eq:modulispaceformula}. It is 
very interesting that in this context, the complexity,
similar to entanglement entropy, can also be computed via a replica trick. To be clear, while the expression \eqref{yN} is calculated in the Euclidean path integral, we have not explicitly shown the existence of replicated geometries. Perhaps one should take the validity of \eqref{yN} as an indication on the existence of some kind of broader approach involving replica geometries. It is also worth noting that the above expression found by use of a replica trick is identical to 
the expression of the matter two-point function \eqref{eq:quantumgravitydisk2pf} with the 
identification of $\Delta=N$. However, although they are the same expression, conceptually 
they play different roles as \eqref{yN} is used in \eqref{eq:replica-formula}. This is where our approach deviates significantly from 
\cite{Iliesiu:2021ari}. 

Indeed, it is not clear if one could interpret \eqref{yN} as a matter two-point function since the 
corresponding matter two-point function is obtained from an opposite limit, namely, in 
the limit of large scaling  
dimension. On the contrary, in our case, we need the limit, $N\rightarrow 0$ by which we lose
the semiclassical interpretation of the two-point function. Nonetheless, as long as 
the computations are concerned, both yield the same result. 

In particular from \eqref{eq:quantumgravitydisk2pf} by use of \eqref{eq:modulispaceformula} 
one gets
\be
\label{eq:y2Nreplica}
\langle y^{2N}\rangle_u=
\frac{e^{S_0}}{Z_{D,\mu}(\beta)}\int_0^\infty dk_1\, dk_2\; 
e^{-\frac{k_1^2}{2}(\beta-u)-\frac{k_2^2}{2}u}\;
r(k_1)r(k_2)\,\gamma_\mu(k_2)\; {\cal N}(N, k_1,k_2)\,.
\ee
Of course this expression in itself does not yet furnish late time linear growth as \eqref{eq:y2Nreplica} is not the end of the story and needs to be plugged into the replica formula \eqref{eq:replica-formula} and analytically continued to find complexity. Performing the analytic continuation $u=\frac{\beta}{2}+i t$ and using energy variables we arrive at 
\be
\langle y^{2N}\rangle_t=
\frac{e^{S_0}}{Z_{D,\mu}(\beta)}\int_0^\infty dE_1\, dE_2\; 
e^{-\frac{\beta}{2}(E_1+E_2)+i(E_1-E_2)t}\;
{\hat \rho}_D(E_1){\hat \rho}_D(E_2)\,\gamma_\mu(E_2)\; {\cal N}(N, E_1,E_2)\,.
\ee
Now we have to simply plug this equation into the replica formula \eqref{eq:replica-formula}. 
Moreover since we are interested in the behaviour at late times, the main 
contribution should come from the coincident limit, $E_1\rightarrow E_2$. In this limit, using the  
change of variables, 
\be
\label{eq:change-variable}
E=\frac{E_1+E_2}{2},\;\;\;\;\;\;\;\omega=E_1-E_2,
\ee 
one gets
\be
\langle\ell(t)\rangle\sim {\rm const.}-\frac{e^{S_0}}{2\sqrt{2}\pi Z_{D,\mu}(\beta)}\int_0^\infty dE
e^{-\beta E}\sqrt{E} {\hat \rho}_D(E)\gamma_\mu(E)
\int_{-\infty}^{\infty} d\omega\; 
\frac{e^{i\omega t}}{\omega^2}\,,
\ee
which results in the linear growth $\langle\ell(t)\rangle\sim t$.
Of course, one still needs to perform the 
integral over $E$, though we will not do it here. Here our aim was only to show that the linear growth at the disk level could be thought of as the consequence of our replica trick. Performing the calculation of the quenched length on a two-boundary topology and using \eqref{eq:sinekernel} would lead to the results already obtained in \cite{Iliesiu:2021ari} and we will therefore not do this explicitly.


\subsection{The geodesic $L$}\label{sec:geodesicL}

It is straightforward to compute the late time behaviour of the quantum expectation value of the 
length of the geodesic connecting a point on the 
boundary to one on the EOW brane
\be
\label{eq:repl1}
\langle L(u)\rangle=-\frac{1}{Z_{D,\mu}(\beta)}\int_0^\infty
\frac{dz}{z}\;\Psi_D(\beta-u,L)\,\Psi_D(u,L)\,\ln\frac{z}{4}=-\lim_{N\rightarrow 0}
\frac{\langle z^N\rangle_u -1}{N},
\ee
where 

\bea
&&\langle z^N\rangle_u=\frac{1}{Z_{D,\mu}(\beta)}\int_0^\infty
\frac{dz}{z}\;\Psi_D(\beta-u,L)\,\Psi_D(u,L)\,\left(\frac{z}{4}\right)^N\\
&&\;\;\;\;\;\;\;\;\;\;=\frac{e^{S_0}}{{2}Z_{D,\mu}(\beta)}\int_0^\infty dk_1\,dk_2 \; e^{-\frac{k_1^2}{2}(\beta-u) -\frac{k_2^2}{2}u}\,
\gamma_\mu(k_1)\gamma_\mu(k_2)r(k_1)r(k_2)\;{\cal M}(N,k_1,k_2)\,.\nonumber
\eea
Here we have introduced

\be\label{eq:WhitakerintegrationM}
{\cal M}(N,k_1,k_2)=\int_0^\infty
\frac{dz}{z^2}\;W_{-\mu,ik_1}(z)\;W_{-\mu,ik_2}(z)\left(\frac{z}{4}\right)^N\,.
\ee
At this point, one could perform a computation similar to what was done in the case of 
$\langle\ell(u)\rangle$ in the previous section to find the late time behavior 
of $\langle L(u)\rangle$. In general, we would expect to get the same linear growth as before, 
although in this case we will have to deal with the Whittaker functions. 
However, we will postpone this computation for a little while and will first study the higher genus 
corrections to the late time behaviour of complexity. The reason for changing the order of 
computation is as follows.
The computation of complexity as the quantum expectation value of the geodesic length at the 
disk level yields a late time linear growth which keeps growing forever. 
However, on general grounds it is expected that complexity saturates at late times. Therefore 
the disk level computation should not constitute the entire story. It is natural to expect that the inclusion of higher topologies and connected geometries plays an important role. Thus, in order to see the saturation phase, one needs  to compute the quantum expectation of geodesic length taking into account surfaces of higher 
genus \cite{Iliesiu:2021ari}. By making use of the trumpet wavefunctions we have found in section \ref{sec:trumpetwf},
one has
\be
\langle L(u)\rangle =-\frac{1}{Z_{\mu}(\beta)}\int b_1db_1\;b_2db_2\; X(b_1,b_2)
\int_0^\infty \frac{dz}{z}\;\Psi_{T}(\beta-u,b_1,L) \;\Psi_{T}(u,b_2,L)\;
\ln \frac{z}{4}\,,
\ee  
where we have used the notation \eqref{eq:X} again.
In this case we compute the following quantity to be used in the replica formula
\be
\langle z^N\rangle_u=\frac{1}{Z_{\mu}(\beta)}\int b_1db_1\;b_2db_2\; X(b_1,b_2)
\int_0^\infty \frac{dz}{z}\;\Psi_{T}(\beta-u,b_1,L) \;\Psi_{T}(u,b_2,L)\; \left(\frac{z}{4}\right)^{N}\,.
\ee
Using equation \eqref{eq:trumpetwavefunctionwitheowandgeodesic} and expression 
\eqref{eq:WhitakerintegrationM} one finds
\bea
\langle z^N\rangle_u\!&\!=\!&\!\frac{2e^{-S_0}}{{\pi^2}Z_{\mu}(\beta)}\int_0^\infty dk_1\,dk_2\;
e^{-\frac{ k_1^2}
{2}(\beta-u)-\frac{k_2^2}{2}u}\,
\gamma_\mu(k_1)\;\gamma_\mu(k_2)\\ &&\;\;\;\;\;\;\;\;\;\;\;\;
\times \int b_1db_1\;b_2db_2\; X(b_1,b_2)
{\cos(k_1 b_1)\cos(k_2 b_2)}{\cal M}(N,k_1,k_2)\nonumber\,,
\eea
which in the energy variable may be reexpressed as
\be
\label{zn-u}
\langle z^N\rangle_u=\frac{2 e^{-S_0}}{Z_{\mu}(\beta)}\int_0^\infty dE_1\,dE_2\;
e^{-E_1(\beta-u)-E_2u}\,
\gamma_\mu(E_1)\;\gamma_\mu(E_2)\;\langle\rho(E_1)\rho(E_2)\rangle{\cal M}(N,E_1,E_2)\,,
\ee
$\langle\rho(E_1)\rho(E_2)\rangle$ being the spectral correlation.
The main part of the above equation is ${\cal M}(N,E_1,E_2)$ which is an integral involving 
Whittaker functions.
This can be evaluated using the integral identity \cite{Gradshteyn2014}
\begin{align}
\label{eq:identity-general-mu}
\int_0^\infty x^{\rho - 1} & W_{k,m}(x) W_{\lambda ,n}(x) = \frac{\Gamma (2 n) \, \Gamma (-m-n+
\rho +1) \, \Gamma (m-n+\rho +1)}{\Gamma
   \left(n-\lambda +\frac{1}{2}\right) \Gamma \left(-k-n+\rho
   +\frac{3}{2}\right)} \nonumber \\ &_3F_2\left(-n-\lambda
   +\frac{1}{2},-m-n+\rho +1,m-n+\rho +1;1-2 n,-k-n+\rho +\frac{3}{2};1\right) \nonumber \\
   &+\frac{\Gamma (-2 n) \, \Gamma (-m+n+\rho +1) \, \Gamma (m+n+\rho +1)
  }{\Gamma \left(-n-\lambda +\frac{1}{2}\right) \Gamma
   \left(-k+n+\rho +\frac{3}{2}\right)}  \nonumber \\ &_3F_2\left(n-\lambda +\frac{1}{2},-m+n+\rho 
   +1,m+n+\rho +1;2 n+1,-k+n+\rho
   +\frac{3}{2};1\right).
\end{align}

In order to evaluate the late time behaviour of complexity, the scheme is as follows. 
First we need to make the analytic continuation $u=\frac{\beta}{2}+it$ as before. Then plugging 
the resulting expression in the replica formula and taking the $N\rightarrow 0$, limit one can find the 
quantum expectation value of the geodesic length or equivalently, the complexity. Now since we 
are only interested in late time behaviour, the main contribution comes from the coincident limit, 
namely, $E_1\rightarrow E_2$. It is convenient to use $E$ and $\omega$ variables as defined in 
\eqref{eq:change-variable}. Using \eqref{eq:identity-general-mu} for our case, we get a nice 
expansion of the function ${\cal M}(N,E_1,E_2)$ in the limit $\omega \rightarrow 0$
\be
\label{eq:dMdN-mu}
\lim_{N\rightarrow 0}\frac{d}{dN} {\cal M}(N,E_1,E_2)=\frac{ \sqrt{2 E} }{2\pi \gamma_\mu(E)
{\hat \rho}_D(E)} \frac{1}{\omega ^2}+{\rm 
local\,
terms}\,.
\ee
However, in order to obtain the late time behaviour of complexity, we still need to perform the 
integrations over $E$ and $\omega$.
Using this and the replica trick detailed above,
one arrives at\footnote{Since we are interested in the time dependence of complexity, 
in this expression we have dropped a local term leading to a
time independent term in the complexity and added all terms into the constant term. 
The corresponding term is divergent and has the form of $\delta(\omega)/\omega$. }
\bea
\label{eq:Lexpectation}
&&\langle L(t)\rangle={\rm const.}-\frac{ e^{S_0}}{\pi Z_{\mu}(\beta)}\int_0^\infty dE\,
e^{-\beta E}\,
\sqrt{2E}\gamma_\mu(E){\hat \rho}_D(E)\\ &&\hspace*{7cm}\times 
\int_{-\infty}^\infty d\omega\,
\frac{e^{i\omega t}}{\omega^2}\,
\left(1
-\frac{\sin^2\left(\pi {\hat \rho}_D(E) e^{S_0}\omega \right)}
{(\pi {\hat \rho}_D(E)e^{S_0}\omega)^2}\right)\,.\nonumber
\eea

It is worth stressing here that in order to derive the expression given in \eqref{eq:Lexpectation}, 
one needs to take into account the non-perturbative effects explicitly through the sine-kernel 
appearing in the spectral correlation given in \eqref{eq:sinekernel} \cite{Saad:2019lba}.

It is now clear that the $\omega$-integral may be performed exactly. In particular 
the expression in brackets on the right hand side of \eqref{eq:Lexpectation} corresponds to the 
disk contribution that results in linear 
growth. As was observed in \cite{Iliesiu:2021ari}, the disk 
linear growth is cancelled by the non-perturbative term as long as $2 \pi {\hat \rho}_D(E)e^{S_0}\ll t$. 
It is easy to check that in this regime the integral vanishes identically. 

On the other hand for $2 \pi {\hat \rho}_D(E)e^{S_0}\gg  t$, expanding the ``sin''-contribution in terms of 
exponentials and deforming the pole one finds \cite{Iliesiu:2021ari}
\be
\int_{-\infty}^\infty d\omega\,
\frac{e^{i\omega t}}{\omega^2}\,
\left(1
-\frac{\sin^2\left(\pi {\hat \rho}_D(E) e^{S_0}\omega \right)}
{(\pi {\hat \rho}_D(E)e^{S_0}\omega)^2}\right)=
\frac{2\pi^2{\hat \rho}_D(E) e^{S_0}}{3 }\, \left(1-\frac{t}{2 \pi  {\hat \rho}_D(E)  e^{S_0}}\right)^3\,.
\ee
Therefore overall
\be
\langle L(t)\rangle={\rm const.}-\frac{ 2\pi e^{2S_0}}{3 Z_\mu(\beta)}\int_{E_0}^\infty dE\,
e^{-\beta E}\,
\sqrt{2E}\gamma_\mu(E){\hat \rho}^2_D(E)\, 
\left(1-\frac{t}{2 \pi  {\hat \rho}_D(E)  e^{S_0}}\right)^3\,.
\ee 
Here $E_0$ is implicitly  obtained via the equation  $\pi {\hat \rho}_D(E_0)e^{S_0}= t$.
Finally we have to perform the integral over $E$. 
To proceed, it is instructive to consider particular values of $\mu$ for which the above 
expression is simplified further. In what follows we will consider the case of $\mu = \frac{1}{2}$ as 
an illustrative example. In this case  using the fact that 
\be
\gamma_{\frac{1}{2}}(E)=\frac{\pi \sqrt{2E}}{\sinh(\pi  \sqrt{{2E}})}\,,
\ee
one gets
\be
\label{L-expectation-2}
\langle L(t)\rangle={\rm const.}-\frac{4\pi^2 e^{2S_0}}{3Z_{\frac{1}{2}}(\beta)}\int_{E_0}^\infty dE\,
e^{-\beta E}\,
 \frac{ E \,{\hat \rho}_D^2(E)}{\sinh(\pi  \sqrt{{2E}})} \left(1-\frac{t}{2 \pi  {\hat \rho}_D(E)  
 e^{S_0}}\right)^3\,,
\ee 
where
\be
Z_{\frac{1}{2}}(\beta)=e^{S_0}\int_0^\infty dE\, e^{-\beta E}\gamma_{\frac{1}{2}}(E)
{\hat \rho}_D(E)=\frac{e^{\pi^2/2\beta}e^{S_0}}
{\sqrt{2\pi}\beta^{3/2}}\left(1+\frac{\pi^2}{\beta}\right)\,.
\ee
For times $t\ll e^{S_0}$ one may expand the r.h.s of \eqref{L-expectation-2} and evaluate the 
integral which at leading order takes the form
\be
\langle L(t)\rangle\approx {\rm const.}-C_0 e^{S_0}+C_1 \,t\,,
\ee
where
\bea
C_0&=&\frac{\pi^2+3 \beta +9 e^{\frac{4 \pi ^2}{\beta }} \left(\beta +3 \pi ^2\right) }{6 \beta  
\left(\beta +\pi ^2\right)}\,,\cr &&\cr
C_1&=&\frac{ \sqrt{2} e^{-\frac{\pi ^2}{2 \beta }} \sqrt{\beta } \left(2 \beta +\pi ^2\right)+\pi ^{3/2} 
\left(3 \beta +\pi ^2\right) \text{erf}\left(\frac{\pi }{\sqrt{2} \sqrt{\beta }}\right)}{\sqrt{\pi } \beta  
\left(\beta +\pi ^2\right)}\,.
\eea
For large $t$ ( $t\sim e^{S}$) the lower limit of the integral becomes large as well;
 $E_0\rightarrow\infty$. Taking into account that the integrand itself has a factor of 
 $e^{-\beta E}$ results in the 
fact that the integral decays and therefore the quantum expectation value of the geodesic length  
becomes constant. This can be interpreted as the saturation of complexity. For large $t$, one can estimate the 
rate by which the integral decays. For large $t$ the lower limit of integral reads
$E_0= \frac{1}{8\pi^2}\ln^2(2\pi e^{-S_0} t)$. In this limit, approximating the ``sinh'' by an
exponential function one arrives at
\be
\langle L(t)\rangle\approx{\rm const.}-\frac{2\beta^{3/2}  e^{-\frac{\pi ^2}{2 \beta }}}{3 \pi^2  
\left(\beta +\pi ^2\right)}\,e^{S_0}\,e^{-\frac{\beta}{8\pi^2}\ln^2(2\pi e^{-S_0} t)}\,
\left(e^{-S_0} t\right)^{3/2}\,\ln^2(2\pi e^{-S_0} t)\,.
\ee

To summarise, our computation shows that the complexity grows linearly at late times up to 
$t\sim e^{S_0}$ and then saturates to a constant value of order $e^{S_0}$.  
Although we have demonstrated this behaviour explicitly only for a particular value of $\mu$, the qualitative late time behaviour of complexity is the 
same for any value of $\mu$. 


\subsection{The variance of complexity}\label{sec:variance}

Although the results of section \ref{sec:complexity} and the results of \cite{Iliesiu:2021ari} exhibit late 
time behaviour in line with general expectations for complexity, this can be probed further by 
calculating the variance $\sigma$. 
Based on the procedure of computing the complexity in terms of the boundary-to-boundary two-point function, the variance of complexity has been evaluated in \cite{Iliesiu:2021ari} where it was 
observed that the fluctuations exhibit linear growth at late times that 
is in tension with general expectations. In particular, this becomes especially problematic as the ``noise" grows to the same size as the ``signal" at $t \sim e^{2 S_0}$. 

Here we would like to use our approach based on the replica trick to compute the variance. To proceed let us focus on the two-sided case first to draw a direct comparison.
Its generalisation to the one-sided case is then evident.


The variance has the structure
\be\label{eq:variancetwosided}
\sigma_{\ell}^2=\langle\ell^2(u)\rangle-
\langle\ell(u)\rangle^2=\langle\ell^2(u)\rangle_{\text{C}}\,,
\ee
where we denote the connected contribution by \text{C}.
Now in line with the rest of this section, it is clear that the quantity we have to determine is
\be\label{eq:connectedvariance}
\langle \ell^2(u)\rangle_{\text{C}}= \frac{4}{Z( \beta)}\int_{0}^{\infty}db_1 b_1 db_2 b_2 
X(b_1,b_2)\int_{0}^{\infty} \frac{dy}{y} \Phi_{T}(\beta-u,b_1,\ell)\Phi_{T}(u,b_2,\ell) \left(\ln \frac{y}
{4}\right)^2\,.
\ee
In order to calculate this we have to apply a replica type formula. We utilise the simple relation
\be \label{eq:replicarelation}
\ln^2 A= \lim_{N\rightarrow 0}\frac{d^2}{dN^2}\; A^N\,.
\ee
By which the  equation \eqref{eq:connectedvariance} may be recast into the following form
\bea
&&\langle\ell^2(u)\rangle_{\text{C}}=\frac{1}{Z(\beta)}\lim_{N\rightarrow 0}\frac{d^2}{dN^2} 
\int_0^\infty b_1db_2\;b_2db_2 X(b_1,b_2) \\ &&\hspace*{6cm}\times\int_0^\infty\frac{dy}{y}
\Phi_T(\beta-u,b_1,\ell)\Phi_T(u,b_2,\ell)\left(\frac{y}{4}\right)^{2 N}\,.\nonumber
\eea
This of course has a structure similar to the calculations of sections \ref{sec:labell} and 
\ref{sec:geodesicL} and it is therefore clear that by making use of the trumpet wavefunction 
\eqref{eq:trumpetwavefunction} one arrives at
\be
\langle \ell^2(u)\rangle_{\text{C}}=\frac{4e^{-S_0}}{Z(\beta)}\int_0^\infty dE_1\;dE_2 
e^{-E_1(\beta-u)-E_2u}\langle \rho(E_1)\rho(E_2)\rangle
\left(\lim_{N\rightarrow 0}\frac{d^2}{dN^2}{\cal N}(N,E_1,E_2)\right)\,,
\ee
which we analytically continue to 
\be
\langle \ell^2(t)\rangle_{\text{C}}
=\frac{4e^{-S_0}}{Z(\beta)}\int_0^\infty dE \int_{-\infty}^\infty\;d\omega 
e^{-\beta E+i\omega t}\langle \rho(E+\frac{\omega}{2})\rho(E-\frac{\omega}{2})\rangle
\left(\lim_{N\rightarrow 0}\frac{d^2}{dN^2}{\cal N}(N,E,\omega)\right)\,,
\ee
where we are using the coordinates \eqref{eq:change-variable}.
At late times, taking the limit $\omega \rightarrow 0$, we have
\be
\lim_{N\rightarrow 0}\frac{d^2}{dN^2}{\cal N}(N,E,\omega)=\frac{\sqrt{E}}{8\pi {\hat \rho}_D(E)}
\;\left(\psi(2i\sqrt{2E})+\psi(-2i\sqrt{2E})-\ln 4\right)\;\frac{1}{\omega^2}+{\cal O}(\omega^0)\,,
\ee
where we have introduced the Polygamma function $\psi(x)$. This may then be used together 
with \eqref{eq:sinekernel} to arrive at the final result
\bea
\label{eq:variancesquared}
&&\langle \ell^2(t)\rangle_{\text{C}}
=\frac{e^{S_0}}{2\pi Z_D(\beta)}\int_0^\infty dE e^{-\beta E}(\psi(2i\sqrt{2E})+\psi(-2i\sqrt{2E})-
\ln 4) {\hat \rho}_D(E)\sqrt{E}\cr &&\cr &&\;\;\;\;\;\;\;\;\;\;\;\;\;\;\;\;\;\;\;\;\;\;\;\;\;\;\;\;\;\;\;\;\;\times
\int_{-\infty}^\infty d\omega
\frac{e^{i\omega t}}{\omega^2} \left(1-\frac{\sin^2(\pi {\hat \rho}_D(E)e^{S_0}\omega)}{
(\pi {\hat \rho}_D(E)e^{S_0}\omega)^2}\right)\,.
\eea
We can see that the $\omega$ integration is of the same form as the one which appears in the calculation of the 
complexity itself.\footnote{Note that in this expression we have not considered a contact term that 
is proportional to a delta function. As we mentioned in the calculation of complexity, this term being 
of the form of $\delta(\omega)/\omega$ leads to a time independent term which does not 
contribute to complexity growth. In the present case this term gives a divergent term which could be 
removed by subtracting $\ell(0)$. Although it is important to consider this term in the computation of variance, since our aim was to show how the replica trick results 
in a reasonable variance, we have just considered $\ell^2(t)$ and dropped the 
corresponding term by hand.} 
Indeed the only difference is the additional Polygamma structure. This is a 
pleasing result. The expression \eqref{eq:variancesquared} circumvents the problematic late time 
growth of noise observed in \cite{Iliesiu:2021ari}. The result saturates to a constant value and we 
therefore recover time-independent fluctuations before the recurrence time. We also observe that \eqref{eq:variancesquared} implies a signal-to-noise ratio of order $\sim e^{- 
\frac{S_0}{2}}$ at $t \sim e^{S_0}$.

For the one-sided black hole the procedure is the same and indeed we recover a similar 
expression with a rather more complicated $E$-dependent function that comes from the fact that
\be
\lim_{N\rightarrow 0}\frac{d^2}{dN^2}{\cal M}(N,E,\omega)=\frac{ 1}{
{\hat \rho}_D(E)}\frac{F(E,\mu)}{\omega^2}+{\cal O}(\omega^0)\,,
\ee
where we introduced $F(E,\mu)$, which is a complicated function of $E$ and $\mu$ containing hypergeometric and polygamma functions and their derivatives.

\section{Conclusion and Outlook}

\label{sec:conclusion-outlook}
In this work, we have considered a fixed EOW brane which plays the role of a cutoff by removing
a part of boundary. This setup provides a holographic model for a one-sided black hole. We have 
computed the multi-boundary partition functions and the matter correlation functions in this model. 
However, the most important result in this work is the computation of complexity. 

To compute complexity we have employed a modified version of the well-known 
replica trick used to study the quenched free energy. This avoids the ambiguity of defining complexity in terms of boundary-to-boundary correlation functions as advocated for in \cite{Iliesiu:2021ari}. The tension between the limit of scaling dimensions and the geodesic approximation is therefore not present in this work.
We have retrieved the expected non-perturbative plateau regime in the late time growth of complexity, which follows an early period of perturbative linear growth in 
time. Although the result is qualitatively similar to that of a two-sided black hole, except for the 
coefficients being sensitive to the tension of the EOW brane now, the replica trick employed in our 
work yields a more satisfactory result for the variance.  The emergence of only time-independent fluctuations in the variance compared to the late-time linear growth of 
\cite{Iliesiu:2021ari} would seem an advancement in the calculation of the black hole volume in JT 
gravity. Of course in our approach the geometric picture is less obvious. 

We will now conclude with a couple of interesting and related questions which are in progress.
\paragraph{Dynamical EOW branes}
So far we have considered a fixed EOW brane without any associated dynamics. However, 
it is interesting to consider a dynamical EOW brane. This requires considering a certain EOW brane 
that contributes to the path integral. In other words, one could imagine a general hypersurface with some of geodesics capped by EOW branes.

\begin{figure}
\begin{center}
\includegraphics[scale=0.9]{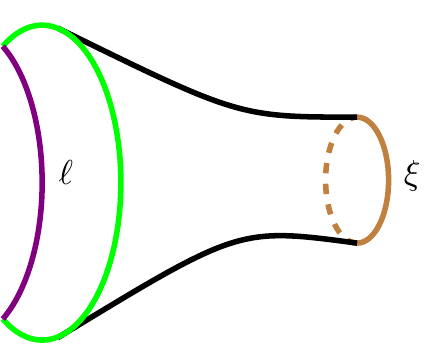}
\end{center}
\caption{Trumpet capped by the FZZT brane shown by a brown circle and parametrized by 
$\xi$. }
\label{fig:FZZT}
\end{figure}

To start with we can start with a toy model where the geodesic of a trumpet geometry is capped 
off by an Fateev-Zamolodchikov-Zamolodchikov-Teschner (FZZT) anti-brane \cite{Fateev:2000ik, Teschner:2000md} as shown in fig.
\ref{fig:FZZT}. Following the prescription of \cite{Okuyama:2021eju}, what we need to do is insert a factor of  $-\frac{1}{b}e^{-\xi b}$ in the path integral on a trumpet with parameter $b$. 

In order to see the effect of this brane on the behaviour of complexity as a function of time, following the 
procedure we adopted for the EOW brane, one first needs to construct the corresponding 
wavefunction in presence of the FZZT anti-brane. In what 
follows, for simplicity, we shall consider two-sided
black holes. Starting from  $\Phi_T(\beta,b,\ell)$ given in \eqref{eq:trumpetwavefunction}, one can 
compute the wavefunction associated with fig. \ref{fig:FZZT} as
 \be
 \label{eq:trumpet-FZZT}
 \Phi_T(\beta, \ell)=  -\int_0^\infty db\,e^{-\xi b}\, \Phi_T(\beta,b,\ell)=-\frac{4e^{-S_0/2}}{\pi}
\int_0^\infty dk\;e^{-\frac{\beta k^2}{2}}\, \frac{\xi}{\xi^2+k^2} K_{2ik}(y).
\ee

With this result in hand, we need to employ our modified replica method defined through 
\eqref{eq:replica-formula} which yields, at late time, 
\bea
\label{eq:l-FZZT}
 \langle \ell(t)\rangle={\rm Const.}+\frac{1}{\pi^3{\tilde Z}(\beta)} \int_0^\infty \frac{dE}{2E}
\, e^{-\beta E}\, &&\bigg[\pi\xi\,\kappa \frac{2 \sqrt{2E}{\hat \rho}_D(E)}{\xi^2+2E}
- \frac{\kappa^2\xi^2 e^{-S_0}}{(\xi^2+2E)^2}\bigg] \nonumber\\
&& \times \frac{\sqrt{2E}}{{\hat \rho}_D(E)}\,\int_{-\infty}^\infty d\omega\frac{e^{i\omega t}}
{\omega^2},
\eea
where $\kappa$ is the number of FZZT anti-branes.
Since we are only interested in the late time behaviour, we have used the $E$ and $\omega$ 
variables \eqref{eq:change-variable} in the coincident limit, $E_1 \rightarrow E_2$.

 The $\omega$-integral of \eqref{eq:l-FZZT} can be readily performed and yields
 \bea
 \label{eq:complexity-FZZT}
 \langle \ell(t)\rangle={\rm Const.}-\frac{ t}{\pi^2{\tilde Z}(\beta)} \int_0^\infty \frac{dE}{2E}
\, e^{-\beta E}\,\bigg[\pi\xi\,\kappa \frac{2 \sqrt{2E}{\hat \rho}_D(E)}{\xi^2+2E}
- \frac{\kappa^2\xi^2 e^{-S_0}}{(\xi^2+2E)^2}\bigg]
\frac{\sqrt{E}}{{\hat \rho}_D(E)}.
\eea 
From \eqref{eq:complexity-FZZT} it is clear that whether the above contribution results in a 
decreasing or increasing behaviour of complexity at late times depends on the $E$ integral. 
Note that the disk contribution is proportional to $e^{S_0}$ whereas the above contribution is 
given in terms of the number of branes $\kappa$, therefore one might naively expect an 
interesting
competition between $\kappa$ and $e^{S_0}$ that is similar to that of entanglement entropy. We hope to report the final conclusion, both for the two-sided 
and one-sided black hole geometries, soon \cite{inprogress}. We expect this computation to shed 
light on the physical interpretation of the replica procedure we employed to compute complexity.

\paragraph{UV cutoff}
In this paper we discussed EOW branes playing the role of cutoffs. In the Lorentzian version of 
the theory, the cutoff EOW brane lies behind the event horizon of the black hole. In holographic 
theories, there is an interesting correspondence between a UV cutoff near the boundary of AdS 
spacetime and a conformal field theory deformed by a particular irrelevant operator quadratic in the stress-energy tensor \cite{McGough:2016lol, Taylor:2018xcy, Hartman:2018tkw}, 
namely, the $T{\bar T}$ deformation 
\cite{Zamolodchikov:2004ce,Smirnov:2016lqw, Cavaglia:2016oda}. 
The wavefunction technique we used for the EOW brane will also be useful in computing complexity for a $T{\bar T}$-deformed CFT.

The partition function of $T{\bar T}$ deformed JT gravity may be written as 
\cite{Iliesiu:2020zld}
\be
\label{eq:partition-TT}
Z_{D,\lambda}(\beta)=\int_{-\infty}^\infty dE\;e^{-\beta f(E)}{\hat \rho}_D(E),
\ee
where $f(E)=\frac{1-\sqrt{1-8\lambda E}}{4\lambda}$, $\lambda$ is the deformation parameter 
and $E$, the energy of the undeformed theory. Clearly for $\lambda\rightarrow 0$ one finds 
the standard partition function. 

Our aim is to compute the complexity for this deformed version of JT gravity. As mentioned above, we will 
use the wavefunction formalism. To do so, one needs to write down the corresponding disk wave 
function for the deformed theory. Using the formalism developed in \cite{Iliesiu:2020zld} for 
$\lambda < 0$ one can easily find the deformed wavefunction as 
\begin{align}
\label{eq:wf-TT}
\Phi_{D,\lambda}(\beta, \ell) =
4e^{S_0/2}\int_0^\infty dE e^{-\beta f(E)}\,{\hat \rho}_D(E)\, K_{2i\sqrt{2E}}(y)\,.
\end{align}
which exactly reproduces the partition function \eqref{eq:partition-TT}.

Once we have the wavefunction \eqref{eq:wf-TT}, we can once again use the modified replica 
method \eqref{eq:replica-formula} to compute complexity. In the late time limit, using the 
coincident variables \eqref{eq:change-variable}, we obtain
\be
\langle \ell(t)\rangle \sim {\rm const.}-\frac{2 e^{S_0}}{\sqrt{2}\pi Z_\lambda(\beta)}
\int_0^\infty dE e^{- \beta f(E)}\,\sqrt{E}\,{\hat \rho}_D(E)\,\int_{-\infty}^{\infty}
d\omega\;\frac{e^{\frac{it \omega}{\sqrt{1-8\lambda E}}}}{\omega^2}\,.
\ee
The integral over $\omega$ can be performed exactly and we arrive at the following expression at 
late time showing linear growth of complexity, as expected from the disk level computation.
\be
\langle \ell(t)\rangle \sim {\rm const.}+\frac{2 e^{S_0}t}{\sqrt{2} Z_\lambda(\beta)}
\int_0^\infty dE e^{- \beta f(E)}\,\frac{\sqrt{E}\,{\hat \rho}_D(E)}{\sqrt{1-8\lambda E}}\,.
\ee

While obtaining the plateau regime of complexity in this setup can be done straightforwardly by adding 
higher genus contributions as before, it will be interesting to study the saturation of complexity in 
this deformed JT setup in presence of an EOW brane. We postpone this for future work.
\acknowledgments
We would like to thank Moritz Dorband, Johanna Erdmenger and Misha Usatyuk for helpful discussions.

\bibliographystyle{JHEP}
\bibliography{literature}

\end{document}